\pdfoutput=1
\documentclass{amsart}
\usepackage{macros}

\usepackage{natbib}

\usepackage[english]{babel}
\usepackage[utf8]{inputenc}
\usepackage{amsmath}
\usepackage{csquotes}
\usepackage{fbb}

\title{Factorization algebra}
\author{Kevin Costello \and Owen Gwilliam}

\begin{document}

\begin{abstract}
Factorization algebras are local-to-global objects living on manifolds, and they arise naturally in mathematics and physics. Their local structure encompasses examples like associative algebras and vertex algebras; in these examples, their global structure encompasses Hochschild homology and conformal blocks. In the setting of quantum field theory, factorization algebras articulate a minimal set of axioms satisfied by the observables of a theory, and they capture concepts like the operator product and correlation functions. In this survey article for the {\it Encyclopedia of Mathematical Physics}, 2nd Edition, we give the definitions and key examples, compare this approach with other approaches to mathematically formalizing field theory, describe key results, and explain how higher symmetries can be encoded in this framework.
\end{abstract}

\maketitle

\tableofcontents

Factorization algebras offer a flexible framework for describing the observables and symmetries of field theories in physics, 
but they also appear naturally in several areas of mathematics,
notably in topology and representation theory.
In this way there is an interesting transfer of insights between different subjects,
and much of the research around them benefits from this transfer.

Factorization algebras are a \emph{minimal} axiom system for the observables of a quantum field theory (QFT). The idea is to write down the weakest possible axioms that capture the concepts of \emph{operators} (called observables in \cite{CG1, CG2}), \emph{operator products}, and \emph{correlation functions}.

For chiral conformal field theories (CFT), these aspects of a field theory are captured by a vertex algebra; similarly, for non-chiral CFTs, these ideas are described by the conformal bootstrap axioms.  
Factorization algebra aim to capture these concepts for a general quantum field theory,
while imposing the weakest restrictions that can usefully describe this part of QFT.
In contrast with algebraic quantum field theory, we do not demand that observables form an associative algebra. \emph{A priori} it is not obvious why one can assign an associative algebra to a quantum field theory (and indeed one cannot in Euclidean signature). Instead, as discussed in section~\ref{sec:AQFT}, associative algebras appear in Lorentzian signature as a \emph{consequence} of the axioms.   

\section{Basics of factorization algebras and their physical meaning}

Let us now state the axioms of a prefactorization algebra, which are very simple. 

\begin{dfn}
Let $M$ be a topological space (which the reader can take to be Euclidean $\RR^n$).  
A {\em prefactorization algebra} $\cF$ on $M$ with values in vector spaces is the following data:
\begin{itemize}
\item There is a (topological)  vector space $\cF(U)$ for each open set $U \subset M$. 
\item  There is a linear map $m_V^U: \cF(U) \rightarrow \cF(V)$ for each inclusion $U \subset V$ of open sets.
\item There is a linear map $m_V^{U_1,\ldots,U_n} : \cF(U_1) \otimes \cdots \otimes \cF(U_n) \rightarrow \cF(V)$ for every finite collection of open sets where each $U_i \subset V$ and where the $U_i$ are pairwise disjoint. The following picture represents the situation.
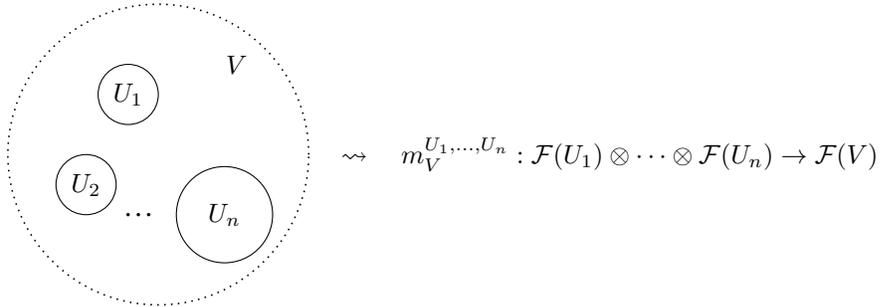
\begin{center}
 \begin{tikzpicture}[scale=0.8]
 \draw[dotted,semithick] (0,0) circle (2.5);
 \draw (-0.5,1) circle(0.5) node {$U_1$};
 \draw (-1.2,-0.5) circle (0.5) node {$U_2$};
 \draw (-0.3, -1) node {\dots};
 \draw (1.1,-1) circle (0.8) node {$U_n$};
 \draw (1.3, 1.5) node {$V$};
\node at (7.5,0){$\rightsquigarrow \quad m_V^{U_1,\ldots,U_n}: \cF(U_1)\otimes\cdots\otimes\cF(U_n)\to\cF(V)$};
 \end{tikzpicture}
\end{center}

\item The maps are compatible in the obvious way, so that if $U_{i,1}\sqcup\cdots\sqcup U_{i,n_i}\subseteq V_i$ and $V_1\sqcup\cdots\sqcup V_k\subseteq W$, the following diagram commutes:
\begin{center}
\begin{tikzcd}[column sep=small]
{\bigotimes}^{k}_{i=1}{\bigotimes}^{n_i}_{j=1}\cF(U_j) \arrow{dr} \arrow{rr} &   &{\bigotimes}^k_{i=1}\cF(V_i) \arrow{dl}\\
&\cF(W)  &
\end{tikzcd}.
\end{center}
\end{itemize}
\end{dfn}

For an explicit example of the associativity, consider the following picture.
\begin{center}
\begin{tikzpicture}[scale=0.9]
\draw[semithick,dotted] (0,0) circle (2);
\draw (1.2, 1.2) node {$W$};
\draw [style= dashed,rotate=45] (-0.1,0.85) ellipse (1.5 and 0.9);
\draw (-0.9, 1.1) node {$V_1$};
\draw [style= dashed] (0.9,-0.5) ellipse (.7 and 1);
\draw (0.9, 0) node {$V_2$};
\draw (-1, 0.2) circle (0.6) node {$U_{1,1}$};
\draw (-0.1, 1.1) circle (0.5) node {$U_{1,2}$};
\draw (0.9,-0.9) circle (0.4) node {$U_{2,1}$};
\node at (6.5,0){$\rightsquigarrow
\begin{tikzcd}[column sep=small]
\cF(U_{1,1}) \otimes \cF(U_{1,2}) \otimes \cF(U_{2,1}) \arrow{dr} \arrow{d}  & \\ 
\cF(V_1) \otimes \cF(V_2) \arrow{r} &\cF(W) 
\end{tikzcd}
$};
\end{tikzpicture}\\
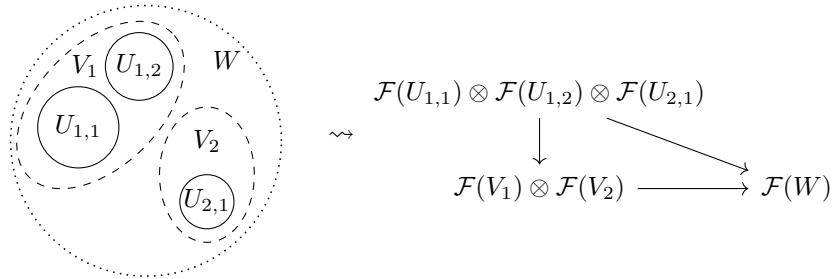
The case of $k=n_1=2$, $n_2 = 1$
\end{center}
Factorization algebras have an additional local-to-global axiom that we discuss in section~\ref{sec:fact}.  

One should think of the vector space $\cF(U)$ as the space of measurements one make in the region~$U$. For instance, in a Lagrangian QFT, $\cF(U)$ will be built from integrals of local operators against test functions supported in $U$, as well as similar multi-local expressions. 

The product on prefactorization algebras is the operator product of QFT. It is only defined on disjoint opens because, in a Euclidean quantum field theory, the operator product expansion has singularities when operators have coincident position.  In a Lorentzian QFT, the singularities can be on the light cone, which suggests a Lorentzian variant of the axioms which we will discuss shortly. 

\subsection{Factorization algebras and associative algebras}
\label{sec:associativealgebras}

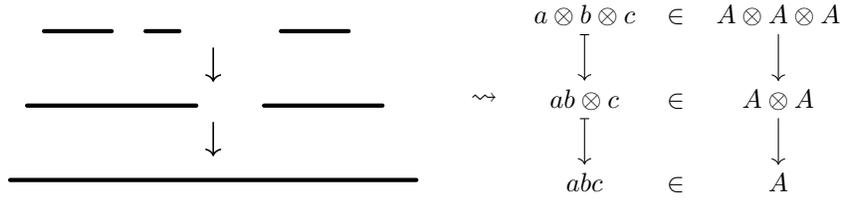
\begin{figure}
\begin{center}
 \begin{tikzpicture}[scale=0.9] 
 \begin{scope}[line cap=round,ultra thick]
 \draw (-2.5,1) -- (-1.5,1); 
 \draw (-1,1) -- (-0.5,1); 
 \draw (1,1) -- (2,1); 
 \draw[->,semithick] (0,0.75) -- (0,0.25);
 \draw (-2.75,-0.1) -- (-0.25,-0.1); 
 \draw (0.75,-0.1) -- (2.5,-0.1);  
 \draw[->,semithick] (0,-0.35) -- (0,-0.85);
 \draw (-3,-1.2) -- (3,-1.2); 
 \end{scope}
 
\node at (4,0) {$\rightsquigarrow$ };

\node at (7,0){
\begin{tikzcd}[cramped,column sep=tiny]
a \otimes b \otimes c \arrow[mapsto]{d} & \in & A \otimes A \otimes A \arrow{d} \\
ab \otimes c \arrow[mapsto]{d} & \in & A \otimes A \arrow{d} \\
abc  & \in &A 
\end{tikzcd}
};
\end{tikzpicture}
\end{center}
\caption{The prefactorization algebra $A^{fact}$ of an associative algebra $A$}
\label{fig:assasfact}
\end{figure}

A crucial example of a factorization algebra is given by an associative algebra. Every associative algebra $A$ defines a prefactorization algebra $A^{\rm fact}$ on the real line $\RR$, as follows. To each open interval $(a,b)$, we set $A^{\rm fact}( (a,b) ) = A$. To any open set $U = \coprod_j I_j$, where each $I_j$ is an open interval, we set $A^{\rm fact}(U) = \bigotimes_j A$. The structure maps simply arise from the multiplication map for $A$. Figure~\ref{fig:assasfact} displays the structure of $A^{\rm fact}$. Notice the resemblance to the notion of an $A_\infty$ algebra, or algebra over the little 1-disks operad. 

Conversely, if a prefactorization algebra $\cF$ on $\RR$ has the property that the structure map $\mc{F}(I) \to \mc{F}(I')$ is an isomorphism for any inclusion of intervals $I \subset I'$,
then $\cF$ gives rise to an associative algebra.  
Moreover, if the prefactorization algebra is equivariant under translation along $\RR$, we find an associative algebra with an infinitesimal automorphism. 
(If it is inner, as in most physical examples, one might call this automorphism the Hamiltonian.)

This class of examples lets us connect with quantum mechanics, viewed as a one-dimensional field theory.
For example, let $A$ be the algebra of operators for a quantum mechanical system. 
We view the associated prefactorization algebra as describing when measurements happen in time: 
during the period $(a,b)$, we have $A^{\rm fact}( (a,b) ) \cong A$ describing possible measurements. 

Similarly, if $\fg$ is a Lie algebra of symmetries of the system,
there is a map of Lie algebras $J: \fg \to A$ and hence a map of associative algebra $J: U\fg \to A$, where $U \fg$ denotes the universal enveloping algebra.
This {\em quantum moment map} determines a map of prefactorization algebras $J^{\rm fact}: U\fg^{\rm fact} \to A^{\rm fact}$.
One views $U\fg^{\rm fact}$ as a current algebra and $J^{\rm fact}$ as encoding how currents determine operators for the quantum system.

\subsection{Correlation functions from prefactorization algebras}

Although the prefactorization algebra axioms are very simple, it is perhaps a little more surprising that this data is enough to capture concepts such as  correlation functions and the operator product expansion. To see this, it is easiest to work on $\RR^n$ and assume that our theory has Euclidean symmetry. 

A \emph{Euclidean-invariant} prefactorization algebra $\cF$ has a compatible action of the Euclidean group $\op{Iso}(\RR^n)$, giving isomorphisms between $\cF(U)$ and $\cF(g (U))$ for any $g \in \op{Iso}(\RR^n)$ in a way respecting products.  

A \emph{state} is a linear map
\[
	\ip{ - } : \mc{F}(\RR^n) \to \CC 
\]
that is invariant under the $\op{Iso}(\RR^n)$ action.  

A \emph{local operator} at $x$ is an element of the inverse limit
\[
	\mc{F}_x = \lim_{r} \mc{F}(D(x,r))  
\]
where $D(x,r)$ is the disc of radius $r$ around $x$. 
Given a local operator $\mc{O}$ at $0$, we can use the translation action to build a local operator $\mc{O}(x)$ at any point~$x$.

Given a state, we define correlation function of local operators, as follows. Given $n$ such operators $\cO_1,\dots,\cO_n$, the correlation function is then defined as 
\[
	\ip{m ( \mc{O}_1(x_1) \dots \mc{O}_n(x_n)   } \in \CC  
\]
where 
\[
	m : \mc{F}_{x_1} \otimes \dots \otimes \mc{F}_{x_n} \to \mc{F}(\RR^n)  
\]
is the factorization product. The limit involved in defining the space of local operators means that $m$ is defined as long as the points $x_i$ are disjoint.  

\subsection{Operator products expansions and prefactorization algebras}

Suppose we have a Euclidean-invariant prefactorization algebra on $\RR^n$.  
Then $\mc{F}(D(0,r))$ is the space of {\em almost-local operators}; 
it is built from products of operators that have been smeared over a disc of radius~$r$. 
The factorization product map
\[
	m : 	\mc{F}(D(0,r)) \otimes \mc{F}(D(x,r) ) \to \mc{F}(D(0,R)),  
\]
where $2 r < \norm{x} < R + r$, has the physical interpretation of the {\em operator product}. 

Wilson's operator product expansion arises as follows. 
First, we take the limit as $r \to 0$, replacing $\mc{F}(D(x,r))$ by $\mc{F}_x$ and $\mc{F}(D(0,r))$ by $\mc{F}_0$.  Then the product is defined for $0 < \norm{x} < R$.  Given two operators $\mc{O}_1$ and $\mc{O}_2$, the product
\[
	m (\mc{O}_1(0), \mc{O}_2(x) ) 
\]
depends smoothly (and, in practice, real-analytically) on $x \in D(0,R))$. 
(Here we are using the fact that $\mc{F}(D(0,R))$ is really a topological vector space.
The smooth dependence here is not a consequence of the axioms of a prefactorization algebra.)  
By construction, for $\norm{x} < \eps$, this product lives in~$\mc{F}(D(0,\eps))$.    

If there exists an asymptotic expansion of $m (\mc{O}_1(0), \mc{O}_2(x) )$  as $\norm{x} \to 0$, then it automatically will live in the space $\mc{F}_0 =\lim_r \mc{F}(D(0,r))$ of local operators. 
Such an asymptotic expansion will be of the form
\[
	m (\mc{O}_1(0), \mc{O}_2(x) ) \sim \sum F_i(x) \mc{O}^i(0)  
\]
where $\mc{O}^i(0) \in \mc{F}_0$ and $F_i(x)$ are real-analytic functions.   
The axioms of a prefactorization algebra do not guarantee the existence of such an asymptotic expansion, but if it does exist, it is unique, and in perturbative QFT this expansion does exist.

The associativity axiom of a prefactorization algebra is intended as a replacement for associativity of the OPE, which for non-conformal theories is very challenging to formulate. 
For chiral theories in dimension $2$, one can show that this OPE satisfies the axioms of a vertex algebra, 
and the associativity axiom of a vertex algebra is a consequence of the associativity axiom of a prefactorization algebra (see~\cite{CG2}).

\section{A history of factorization algebras, and their relationship to other axioms} 

We connect these notions to other approaches and place them in historical context.

\subsection{A brief history}

There are three different notions of factorization algebra in the literature,
and while they share the same spirit, they differ in detail.
The original version is due to \cite{BD}, from the early 1990s,
and it arose as a recasting of vertex algebras (and related notions from chiral conformal field theory) in the language of algebraic geometry;
it has primarily been developed in the setting of complex algebraic curves.
Their theory has profound applications in geometric representation theory, notably the geometric Langlands program,
and its higher-dimensional consequences are still being explored.
The second version began with Francis, Gaitsgory, and Lurie around 2007 (although with roots in earlier work of \cite{SegFA}, \cite{McDuff}, \cite{Salv}, and others).
These authors recognized that a topological analogue should exist to the Beilinson-Drinfeld version,
much as a vertex algebra is reminiscent of an algebra over the little disks operad (i.e., an $E_2$ algebra).
It works in the setting of manifolds (not necessarily smooth), possibly equipped with data like an orientation or a $G$-bundle, but it is not sensitive to more geometric structure (like a metric).
This theory is called {\it topological chiral homology} by \cite{LurieHA} or {\it factorization homology} by \cite{AF1} or {\it blob homology} by \cite{MW}, 
and it has beautiful applications in algebraic topology itself (notably a nonabelian generalization of Poincar\'e duality),
but it also offers a class of computable topological field theories,
a topic we discuss below.
(See \cite{AFprim} for a recent extensive survey. Morrison and Walker came to these ideas by a different route.)
The third version grew in response to the other two, as we recognized around 2008 that the observables produced by BV quantization seemed to form factorization algebras.
In  \cite{CG1,CG2} we developed axioms that fit well with differential geometry and field theory.
All three versions have overlapping domains of applicability,
but comparison is somewhat subtle and technical.
We will focus here on the third version, and its offshoots.

\subsection{Segal's axioms and prefactorization algebras}

The concept of factorization algebra has a close relationship with Segal's axioms for quantum field theory (see \cite{SegalCFT,segal2010locality, segal2014geometric, Kontsevich:2021dmb}).  
Segal  (following work with Atiyah on topological field theories, and ideas suggested by Witten) proposed that a quantum field theory in dimension $d$ is a functor $Z$ that assigns to a closed Riemannian manifold $(N,g)$ of dimension $d-1$, a (topological) vector space $Z(N,g)$, and that assigns to a Riemannian cobordism, a (continuous) linear map.

A quantum field theory in the sense of Segal gives rise to a prefactorization algebra on any Riemannian manifold $M$, as follows. 
Let $Z$ be a Segal QFT. 
If $U \subset M$ is open and $\partial U$ is its boundary, then we set
\[
	\mc{F}_Z(U) = Z (\partial U). 
\]
(To be pedantic, one needs to restrict to opens $U \subset M$ that have a nice boundary.)
If $U_1,\dots,U_n \subset V$ are disjoint opens, 
then the closure $W$ of the manifold $V \setminus (U_1 \cup \dots U_n)$ provides a cobordism between $\partial U_1 \cup \dots \partial U_n$ and $\partial V$, and the factorization product
\[
m: \bigotimes_i \cF_Z(U_i) \to \cF_Z(V)
\] 
is the map~$Z(W)$.

Segal's axioms, however, are much stronger than those of a prefactorization algebra. 
In some ways this is a disadvantage: it is very hard to construct a solution to Segal's axioms, even in perturbation theory. 
By contrast, one can rigorously construct factorization algebras associated to field theories in perturbation theory.  

\subsection{AQFT and factorization algebras}
\label{sec:AQFT}

Algebraic quantum field theory arises from the same impulse as factorization algebras:
one should organize observables by their spacetime support (see \cite{HaagKast, HaagLQT}).
The context and details differ, however, as AQFT focuses on Lorentzian field theories and typically works with $C^*$ or von Neumann algebras.
Moreover, an AQFT only assigns an algebra to a specific class of subspaces or manifolds.

Relationships between the approaches are emerging, especially as modifications and generalizations of traditional AQFT are being actively explored:
\begin{itemize}
\item In \cite{Benini_2019}, Benini, Perin, and Schenkel carefully compare the two frameworks, and they introduce a useful variation --- called {\em time-orderable prefactorization algebras} --- that are well-suited to Lorentzian field theories. Adding the natural hypotheses of Cauchy constancy and additivity, they produce an equivalence between AQFTs and time-orderable prefactorization algebras.
\item \cite{GwiRej1} compare the BV quantization of free theories (such as a free scalar field or a free fermion) in the AQFT and factorization setting, 
showing how they are equivalent and explicating how to transfer information between the two descriptions of observables. In \cite{GwiRej2}, they show that the perturbative AQFT construction for interacting field theories can be extended to produce a factorization algebra of observables.
\item In \cite{benini2022quantization}, Benini, Musanti, and Schenkel  show, using a different framework for quantizing free Lorentzian theories, how to compare their AQFT and factorization algebras.
\end{itemize}
The works of these authors demonstrates one nice feature of factorization algebras: the associative algebras demanded in algebraic quantum field theory arise as a \emph{consequence} of the apparently weaker definition of factorization algebra.

We have already seen this phenonemon in our discussion of quantum mechanics, 
where we saw that prefactorization algebras on $\RR$ (satisfying an extra axiom) give rise to associative algebras, where the associative product comes from the factorization product.

\cite{Benini_2019} shows that something similar happens for prefactorization algebras on a Lorentzian manifold~$M$, if we modify the definition of prefactorization algebras in two ways: 
\begin{enumerate} 
\item[(I)] First, the factorization product is only defined when the opens $U_1,\dots,U_n \subset V$ are {\em time-orderable} (and not just disjoint), 
which means that there is some total ordering $\{\sigma(1) < \sigma(2) < \cdots < \sigma(n)\}$ such that the causal future of an open $U_{\sigma(i)}$ is disjoint from $U_{\sigma(j)}$ if $\sigma(i) < \sigma(j)$. 
Loosely speaking, one can arrange the opens from ``late'' to ``early.'' 
\item[(II)] The {\em time-slice axiom} (or Cauchy constancy) states that if $U\subset V$ are opens in $M$ that are both globally hyberbolic and that share a common Cauchy hypersurface $X \subset U \cap V$, then the map $\mc{F}(U) \to \mc{F}(V)$ is an isomorphism.
This axiom is a form of causality: it says that anything that can be measured in a globally hyperbolic manifold $V$ can be measured in a small neighbourhood of a Cauchy hypersurface.     
\end{enumerate}
In this situation, \cite{Benini_2019} shows that $\mc{F}(U)$ is an associative algebra for any globally hyperbolic submanifold $U \subset M$.  
See their entry \cite{BSEnc} for discussion and context.
We feel that this construction of Benini et al. provides an important conceptual improvement to algebraic quantum field theory. 

\subsection{Chiral conformal field theory and factorization algebras}

Conformal field theory (CFT) in two dimensions, particularly chiral CFT, has experienced extensive exploration by mathematicians as well as physicists,
with deep consequences for representation theory, complex geometry, topology, and operator algebras.
There are now many different mathematical formulations --- including (but not limited to) vertex algebras, conformal nets, and Segal's functors --- and comparisons in various directions.
We list here a few results that entangle factorization algebras, in the style described here, with those other approaches.
\begin{itemize}
\item Chapter 5 of \cite{CG1} examines prefactorization algebras on $\CC$ and identifies properties that allow one to extract a vertex algebra, leading to a functor from this class of prefactorization algebras to the category of vertex algebras. This functor produces many examples, such as the universal affine vertex algebra for an affine Lie algebra $\widehat{L\fg}$, Virasoro vertex algebras (by \cite{WilVir}), and chiral differential operators~(by \cite{GGW}).
\item \cite{Bru} gives a functor from $\ZZ$-graded vertex algebras to factorization algebras on $\CC$. His construction provides a kind of inverse to the functor of \cite{CG1}, and it builds upon work of~\cite{Hua}. 
\item \cite{Hen} shows that a conformal net of finite index is a factorization algebra on the circle. He uses this result in a larger project to understand the 3-category of conformal nets and to construct important examples such as WZW theories and Chern-Simons theories.
\item \cite{HK} provide a functor from locally constant Beilinson-Drinfeld factorization algebras to locally constant Costello-Gwilliam factorization algebras. It would be interesting to extend it to encompass the non-locally constant algebras.
\end{itemize}

\subsection{Topological field theory and factorization algebras}

Topological field theories (TFTs) provide another area where mathematicians have explored deeply, along with physicists.
On the one hand, there is a collection of examples that seem endlessly productive, such as the A- and B-models appearing in mirror symmetry, Chern-Simons and Rozansky-Witten theories related to knot and 3-manifold invariants, and four-dimensional gauge theories related to the Donaldson and Seiberg-Witten invariants.
On the other hand, there is the functorial framework, introduced by  \cite{AtiTFT} and  \cite{SegCFT}, that has evolved into a powerful tool for analyzing TFTs,
particularly nonperturbative aspects rooted in topology (see \cite{Getal, Sharpe, FMT}).
Joining the two hands is a work in progress.

Factorization algebras connect with both directions.
\begin{itemize}
\item In Section 4.1 of \cite{LurieTFT}, Lurie explains why factorization homology should yield a fully extended framed $n$-dimensional TFT from an $E_n$-algebra,
and Scheimbauer gave a detailed development of this idea in her thesis~\cite{Scheim}.
Further structural results by \cite{AFPK,AFT2,MW} enrich this story.
\item This approach has been deployed on concrete examples, notably by Ben-Zvi, Brochier, and Jordan, who computed the factorization homology on surfaces of the representations of a quantum group in \cite{BBJ1,BBJ2}. There are many outgrowths of this work, such as  \cite{cooke2019excision,Brochier_2021,KirTham,brochier2023classification,keller2023finite}.
\item \cite{ES} show how a prefactorization algebra on $\RR^n$ with an action of a super translation algebra (e.g., the observables of a supersymmetric field theory) determines an $E_n$ algebra after choosing a ``twist'' in the super Lie algebra. 
\item There is extensive work on the rigorous construction of perturbative topological field theories, such as Chern-Simons theory or Poisson $\sigma$-models, going back to \cite{AxeSing} and  \cite{KonECM}. Due to the main theorem of \cite{CG2}, such work yields factorization algebras, with the local behavior producing an $E_n$-algebra (see \cite{ES} as well). For some examples, the values of factorization homology have been computed, such as for topological B-models by \cite{LiLi}, Rozansky-Witten theories by \cite{ChanLeungLi}, and the Kapustin-Witten gauge theories by~\cite{EGW}. 
\end{itemize}
A recent expository source is~\cite{Tetal}.

\section{From prefactorization algebras to factorization algebras}
\label{sec:fact}

Now let us get  a bit more formal and discuss the local-to-global axiom that distinguishes factorization algebras from prefactorization algebras.

A factorization algebra is, in essence, a prefactorization algebra whose value on large open sets is determined by its behavior on small open sets.
In other words, its global behavior is governed by its local structure.
We will use terminology from sheaf theory to make the local-to-global condition precise,
but similar intuition motivates ordinary homology in algebraic topology (which is covariant in open sets, just as a prefactorization algebra is).
See Remark~\ref{rmk:AFL} for more on this comparison.

We have already seen how to define correlation functions using prefactorization algebras. 
Let us generalize the discussion and consider a prefactorization algebra on a manifold $M$ instead of on $\RR^n$. 
A state is given by a linear map
\[
\langle \cdots \rangle: \cF(M) \to \CC
\]
that allows us to define correlators of operators in disjoint opens $U_1 \dots, U_n$ via the composition
\[ 
	\cF(U_1) \otimes \dots \otimes \cF(U_n) \to \cF(M) \to \CC. 
\]
In particular, we have $n$-point correlation functions for any $n$-tuple of operators supported at points.

We would like to say that a state is the same data as its collection of correlation functions of operators.   Equivalently, we would like to say that the vector space $\cF(M)$ is the \emph{universal recipient} of correlation functions.  
That is, there are correlation functions valued in $\cF(M)$, 
and any set of correlation functions valued in some other vector space $V$ is obtained via a linear map from $\cF(M)$ to~$V$.

The factorization algebra axiom aims to capture this idea. 
It was inspired by the work of \cite{BD}, \cite{AF1}, and~\cite{LurieHA}. 
Before we give the definition, let us point out that this axiom is tailored to \emph{local} operators.  In general, it does not aim to capture extended defects.   

The key definition that allows us to formulate the local-to-global axiom is the following.

\begin{dfn}
Given a manifold $M$ and an open set $U \subset M$, a {\em Weiss cover} of $U$ is a collection of open subsets $\fV = \{V_i\}_{i \in I}$ with the property that any finite set of points $\{x_1,\ldots,x_n\}$ in $U$ is contained in some $V_i$ from~$\fV$.
\end{dfn}

This notion is stronger than the usual notion of cover, which simply requires that any single point in $U$ is contained in some open set in the cover.  
A typical Weiss cover of $U$ is given by the collection of opens in $U$ that are finite disjoint unions of open balls of radius less than $ \delta$, for some small $\delta$.  
It is a Weiss cover because any finite subset is inside such a union of open balls.  
Typically, Weiss covers contain infinitely many elements, because for a manifold, 
we can always find a finite set not contained in some finite collection of proper open subsets.

\begin{dfn}
A {\em factorization algebra} on $M$ is a prefactorization algebra satisfying
\begin{enumerate} 
\item[(a)] If $U_1$ and $U_2$ are disjoint opens in $M$, then the map $\mc{F}(U_1) \otimes \mc{F}(U_2) \to \mc{F}(U_1 \amalg U_2) $ is an isomorphism, and
\item[(b)] $\mc{F}$ satisfies the cosheaf condition (dual to the sheaf condition) for Weiss covers, 
which means that if the $\{V_i\}$ form a Weiss cover of $U$, then the sequence 
\begin{equation}
\label{eq:cosheaf}
\bigoplus_{i \neq j} \mc{F}(V_i \cap V_j ) \to \bigoplus_i \mc{F}(V_i) \to \mc{F}(U) \to 0  
\end{equation}
is exact on the right and in the middle. 
\end{enumerate}
\end{dfn}

\begin{rmk}
In our book \cite{CG1} we use the {\em derived} version of this condition. 
In that case, each $\mc{F}(U)$ is a cochain complex, 
and we ask that $\mc{F}(U)$ is quasi-isomorphic to the \v{C}ech double complex arising from a Weiss cover. 
In order to make this survey as broadly accessible as possible, we have decided to present the un-derived version of the definition.~\hfill$\Diamond$
\end{rmk}

Let us unpack this definition in the case that we use the Weiss cover built from disjoint unions of open balls of radius less than some small $\delta$. 
Given points $x_1,\dots,x_n \in U$ and radii $\eps_1,\dots,\eps_n < \delta$ such that the balls $D(x_i,\eps_i)$ are pairwise disjoint, 
we have a map  
\[
\bigotimes_i \mc{F}(D(x_i,\eps_i ) ) \to \mc{F}(U). 	
\]
For $\cF$ a factorization algebra, this map satisfies two conditions.
First, the span of the images of these maps, as $x_i$ and $\eps_i$ varies, is all of $\mc{F}(U)$.  
That is, any element of $\mc{F}(U)$ is built from a sum of products of almost-local operators.  
Therefore, any linear functional on $\mc{F}(U)$ is indeed determined by its correlation functions on operators in $\mc{F}(D(x,\eps))$ for $\eps < \delta$ and $\delta$ arbitrarily small. 

The second condition deals with a redundancy in this construction.  
Pick a linear function on $\mc{F}(U)$, giving correlation functions for almost-local operators.  
Pick some operators $\mc{O}_1,\dots,\mc{O}_n$ in $\mc{F}(D(x_i,\eps_i))$.  
Suppose $D(y_i,\delta_i)$ are open discs obtained by moving the discs $D(x_i,\eps_i)$ a little, 
and suppose that $\mc{O}_i$ come from the intersection $\mc{F}( D(y_i,\delta_i) \cap D(x_i,\eps_i) )$.  
Then the correlators of the operators $\mc{O}_i$, viewed as elements of $\mc{F}(D(x_i,\eps_i))$ and as elements of $\mc{F}(D(y_i,\delta_i))$, should be the same.  Asking that the sequence \eqref{eq:cosheaf} is exact in the middle ensures that this is so (and imposes other similar constraints).  

Given a factorization algebra $\cF$ on $M$, 
we often call its global sections $\cF(M)$ (and especially the derived version) the {\em factorization homology} of $\cF$ on $M$.
A common notation is~$\int_M \cF$.

An example might help clarify what information is contained in factorization homology.

\begin{example}\label{ex:HH0}
Let $A$ be an associative algebra.
It determines a factorization algebra $A^{\rm fact}$ on the circle with values in vector spaces, where 
\begin{itemize}
\item for any {\em proper} open set $U = I_1 \sqcup \cdots \sqcup I_k$ consisting of finitely many open intervals, $A^{\rm fact}(U) = A^{\otimes k}$ and
\item $A^{\rm fact}(S^1) = A/[A,A]$, the quotient of $A$ by the vector subspace spanned by commutators $ab - ba$.
\end{itemize}
This quotient space is the source of all trace maps: any linear map $\tau: A \to R$ satisfying $\tau(ab) = \tau(ba)$ for all $a,b\in A$ corresponds to a linear map $A/[A,A] \to R$.
(Compare with the use of density matrices in quantum mechanics.)~\hfill$\Diamond$
\end{example}

This example generalizes substantially. 
We will state a version that is easy to interpret.

\begin{thm}
Let $A$ be a dg algebra. It determines a factorization algebra on $S^1$ assigning $A$ to each open interval inside $S^1$ and whose factorization homology is quasi-isomorphic to the Hochschild complex $C^{\rm Hoch}_*(A,A)$.
\end{thm}

One can thus view factorization homology as providing a systematic generalization of Hochschild homology, with $\int_M \cF$ reflecting the shape of the given manifold~$M$.

\begin{rmk}
\label{rmk:AFL}
A factorization algebra $\cF$ is {\em locally constant} if the structure map $\cF(D) \to \cF(D')$ is a quasi-isomorphism whenever we have an inclusion of disks $D \subset D'$.
In this setting, Ayala and Francis show that the factorization homology satisfy axioms akin to the Eilenberg-Steenrod axioms for homology theories, although the ``coefficients'' (the value on a disk) are now much richer than an abelian group.
(For framed $n$-dimensional manifolds, the coefficients are $E_n$-algebras, i.e., algebras over the little $n$-dimensional disks operad.)
Their characterization offers a powerful toolkit in the setting of algebraic topology,
notably some alternative computational methods.
For a physicist, the observables of a topological field theory provide examples of locally constant prefactorization algebras.~\hfill$\Diamond$
\end{rmk}

\section{Factorization algebras from deformation quantization}
\label{sec:obsasfactalg}

Axiom schemes for quantum field theory are not so useful unless one can produce solutions to these axioms.  In \cite{CG1,CG2} we show that essentially any perturbative field theory in Riemannian signature has a factorization algebra of observables (see Section~\ref{sec:keyresults} for more). 
These books also explore how many standard notions and constructions can be articulated in terms of factorization algebras.
Using these methods, one obtains a direct path between concrete Lagrangian formulations and the structural points of view.
Recent work indicates that analogous results hold for perturbative theories in Lorentzian signature (see~\cite{Benini_2019,GwiRej2}).

In this section we will outline why prefactorization algebras appear in classical field theory (even nonperturbatively) and why perturbative quantization deforms this classical structure. We discuss nonperturbative issues in Section~\ref{sec:nonpert}.

\subsection{Observables as prefactorization algebras}

So far in our discussion of factorization algebras, we have focused on factorization algebras valued in vector spaces.  For a proper discussion of field theory, we need to use factorization algebras valued in cochain complexes.  This is because we work in the BV-BRST formalism, which is needed to quantize gauge theories (and is useful even without gauge symmetry).  The reader uncomfortable with this homological language can generally ignore this extra technology and continue to take the factorization algebras to be valued in vector spaces. 

A classical field theory $\cT$, specified by a Lagrangian density or action functional,
yields a system of partial differential equations that we will call the equations of motion for $\cT$.
In a setting like a gauge theory, we may want to consider solutions to these equations up to an equivalence relation (e.g., gauge transformation) that respects locality.
(It is better, in mathematical practice, to work with a stack of solutions that encodes how different solutions are identified.)
In brief, a classical field theory produces a sheaf $\Sol_\cT$ on spacetime $M$ that assigns to each open set $U$, a space $\Sol_\cT(U)$ of solutions to the equations of motion on $U$.
(This space might be a manifold, possibly infinite-dimensional, or a stack or a derived stack,
depending on the context.)
It is a sheaf because PDE impose constraints that are local in spacetime, and a solution is determined by its behavior on a cover.

Now consider the observables of this theory, which should describe functions on solutions.
Let $\Obs_\cT$ assign to each open set $U$, the commutative algebra of functions on $\Sol_\cT(U)$.
If $\Sol_\cT(U)$ is a (possibly derived) stack, then this algebra $\cO(\Sol_\cT(U))$ will be a dg commutative algebra.
Let $\cO$ denote the functor that takes a space $X$ (e.g., a derived stack for some theories) and produces its algebra of functions $\cO(X)$ (e.g., a dg commutative algebra for those theories). 
This functor $\Obs_\cT = \cO \circ \Sol_\cT: \Open(M) \to \CAlg$ is covariant, because $\cO$ is contravariant and $\Sol_\cT$ is also contravariant in open sets.

Note that this functor $\Obs_\cT$ provides a prefactorization algebra because
\[
\Sol_\cT(U \sqcup V) \cong \Sol_\cT(U) \times \Sol_\cT(V)
\]
for disjoint open sets $U, V$, and so there is a canonical map 
\[
\cO(\Sol_\cT(U)) \otimes \cO(\Sol_\cT(V)) 
\to 
\cO(\Sol_\cT(U) \times \Sol_\cT(V)) 
\]
that gives the structure map
\[
\Obs_\cT(U) \otimes \Obs_\cT(V) \to \Obs_\cT(U \sqcup V).
\]
This map just says that observables supported on disjoint regions can be multiplied.
All the other structure maps can be built from this one and the maps $\Obs_\cT(U) \to \Obs_\cT(U')$ arising from an inclusion $U \subset U'$.

This argument is robust, and it does not depend on the signature of the metric or other details of the theory, just the locality of classical Lagrangian field theory.
It applies nonperturbatively.

Two separate questions now arise.
First, under what conditions is $\Obs_\cT$ a factorization algebra? 
And second, what happens when we quantize $\cT$?

The second question depends on what notion of quantization is used.
So far, research has focused on BV quantization, 
as it produces a deformation of the differential on the observables and hence is well-suited to this setting.
Other methods of quantization might also apply,
but they have not yet been explored in this context,
so far as we know.
Hence we will discuss only BV quantization here.

Even at the classical level, one does not expect the observables of an arbitrary classical field theory to form a factorization algebra.  
Physically, it is because the local-to-global axiom is linked to local operators, and classical gauge theories can have extended defects such as Wilson lines.   
Mathematically, it is because the functor $\cO$ that takes a space $X$ and produces its algebra of functions $\cO(X)$ can lose information; for example, functions on a quotient stack like $BG = */G$ are trivial but the space is not.
It is thus unlikely that the composite $\Obs_\cT = \cO \circ \Sol_\cT$ is a factorization algebra.
(The observables of a Dijkgraaf-Witten theory are typically not a factorization algebra.)

\begin{rmk}
One takeaway is that prefactorization algebras appear naturally, 
but we do not yet know what local-to-global axiom is appropriate in nonperturbative settings.
For TFTs, variants are known: \cite{MW} introduced blob homology,
and \cite{AFR} have enhanced factorization homology,
in steps toward a potential proof of the cobordism hypothesis described by~\cite{AFcob}.~\hfill$\Diamond$
\end{rmk}

In the perturbative setting, however, the situation improves considerably.
By a perturbative classical field theory,
we mean that we fix a global solution $\phi \in \Sol_\cT(M)$ and we study formal deformations of this solution.
In this case, there is a robust mathematical framework, known as derived deformation theory and compatible with the BV-BRST formalism,
and a key result is that for a formal moduli space, its algebra of functions knows everything (if one remembers a canonical filtration).
Since the perturbative space $\Sol_{\cT,\phi}$ of solutions is a sheaf and hence satisfies a local-to-global condition,
the observables then also satisfy a local-to-global condition.
In fact, these classical observables form a factorization algebra.

If a BV quantization exists (i.e., the anomaly vanishes), it produces, in essence, a deformation of the differential that produces a deformation of $\Obs_\cT$ as a factorization algebra.
Such a BV quantization involves issues of renormalization and homological algebra,
and it is these issues that are the focus of \cite{CG2, FR, GwiRej2}, and~\cite{Benini_2019}.

\subsection{Key results for perturbative theories}
\label{sec:keyresults}

In this section, whenever we speak of a perturbative classical or quantum field theory, we will mean the definitions articulated in \cite{CosBook,CG1, CG2}, which encompass the Euclidean versions of many field theories studied in physics and mathematics. The main result is the following.

\begin{thm}
\label{thm:main}
The observables of a perturbative classical field theory on $M$ form a factorization algebra $\Obs^{cl}$ that assigns to every open set, a dg commutative algebra. 
A Batalin-Vilkovisky quantization of this theory, if it exists, yields a factorization algebra $\Obs^q$ that is a flat deformation of $\Obs^{cl}$ over~$\RR[[\hbar]]$.
\end{thm}

This theorem provides an elegant interpretation of Batalin-Vilkovisky formalism as a kind of deformation quantization:
the classical world is commutative and the quantum world deforms it in a noncommutative direction. 
An important refinement, which we will not examine here, is that the classical observables have a 1-shifted Poisson structure (the BV anti-bracket), which is the analogue of the usual, unshifted Poisson bracket in traditional deformation quantization.

\begin{rmk}
One might hope that a similar statement holds for Lorentzian theories as well, and recent work (discussed in Section~\ref{sec:AQFT}) gives cause for optimism.~\hfill$\Diamond$
\end{rmk}

Symmetries play a key role in analyzing and organizing field theories,
and they can be encoded using factorization algebras.
We saw in section~\ref{sec:associativealgebras} how a Lie algebra $\fg$ acting on an associative algebra $A$ by inner derivations (i.e., given a Lie algebra map $\fg \to A$) determines an associative algebra map $U \fg \to A$ and hence a map of factorization algebras $U\fg^{\rm fact} \to A^{\rm fact}$.
This pattern continues. 
Any Lie algebra $\fg$ (or $L_\infty$ algebra) has an {\em enveloping factorization algebra} $\UU \fg$ on a manifold $M$ (see \cite{Knudsen} for a thorough treatment.)
More generally, any reasonable sheaf $\cG$ of Lie algebras on a manifold $M$ has an {\em enveloping factorization algebra} $\UU \cG$ (see chapter 3, section 6 of \cite{CG1}, where $\cG$ is a {\em local $L_\infty$ algebra}).
A rich source of examples are the current algebras of physics,
and these allow for a systematic extension of constructions from mechanics (like quantum moment maps) and conformal field theory (like the Kac-Moody or Virasoro vertex algebras).

An interesting feature of this construction is that $\UU \cG$ has a natural filtration and its associated graded is a 1-shifted Poisson algebra.
In other words, $\UU \cG$ is itself a kind of BV quantization,
much as the universal enveloping algebra $U\fg$ is a deformation quantization of $\Sym(\fg) = \cO(\fg^*)$.
This feature leads to the following refinement of Noether's (first) theorem in this setting for field theory.
See the final chapters of \cite{CG2} for a full discussion and many examples.

\begin{thm}
\label{thm pert Noether}
Let a local $L_\infty$ algebra $\cG$ act as symmetries of a classical field theory on $M$; in particular, it acts by Hamiltonian vector fields for shifted Poisson structure on $\Obs^{cl}$.
Then there is a map of factorization algebras
\[
J^{cl}: \Sym(\cG_c[1]) \to \Obs^{cl}
\]
that respects shifted Poisson structures.

If the classical theory admits a BV quantization, then there may be a cohomological obstruction $\alpha$ to deforming $J^{cl}$ to a map of factorization algebras $\UU \cG \to \Obs^q$.
This cocycle determines a central extension of $\cG$ as an $L_\infty$-algebra, and hence there is a map of factorization algebras 
\[
J^q: \UU_\alpha \cG \to \Obs^q
\]
where $\UU_\alpha \cG$ is the $\alpha$-twisted enveloping factorization algebra of~$\cG$.
\end{thm}

Because we use sheaves of $L_\infty$-algebras, 
our formulation of Noether's theorem incorporates the infinitesimal higher symmetries of \cite{Getal} as well as ordinary symmetries.

The central extension is called a {\em 't Hooft anomaly} in physics literature.
(The term ``anomaly'' is a bit overloaded in physics. 
It means two quite distinct things: a gauge anomaly is an obstruction to quantizing a system, and a 't Hooft anomaly is a central extension of a symmetry algebra. 
The mathematical terminology, of obstruction and central extension, is typically clearer.)   

This formalism, combining BV quantization, factorization algebras, and Noether's theorem, has been deployed in a number of nontrivial applications:
\begin{itemize}
\item \cite{GLL} give a direct connection between Fedosov quantization and the factorization algebra of a 1-dimensional field theory, obtaining a new proof of the algebraic index theorem of \cite{NT1, NT2} (a cousin of the Atiyah-Singer index theorem). See \cite{GuiLi1} for further progress.
\item \cite{GGW} constructed rigorously a 2-dimensional QFT depending on a Calabi-Yau manifold $X$, known as the curved $\beta\gamma$ system, and showed the factorization algebra recovers the chiral differential operators of~$X$, verifying a conjecture of Witten and Nekrasov. Subsequently, \cite{GuiLi2} used this theory to produce a chiral algebraic index theorem.
\item \cite{LiLi} constructed the topological B-model and show the factorization algebra recovers the expected topological vertex algebra (see \cite{LiVOA} for a deeper treatment).
This provides a direct connection to mirror symmetry, a topic of active interest in mathematics for the last three decades.
\item \cite{RabAxial} showed how the local index theorem of Getzler and the axial anomaly are simultaneously captured by quantizing free fermions. 
Subsequently, he recovered the Quillen determinant line in~\cite{RabDet}.
\item \cite{CosYangian} introduced a 4-dimensional gauge theory whose factorization algebra recovers the Yangian, a quantum group introduced by Drinfeld,
and he subsequently gave extensive applications to quantum integrable systems with Witten and Yamazaki in~\cite{CWY1,CWY2, CY}.
\item Costello, Gaiotto, Li, Paquette, and others have explored the AdS/CFT correspondence and its interaction with topological string theory. See ~\cite{CosGaiHol, BudGai, CosPaq}, and \cite{CosLiAnom} as starting places.
\end{itemize}
Much of the work by the community that uses this formalism is focused on holomorphic or topological field theories, primarily because of the researchers' backgrounds, not because the techniques only apply to such theories.

\section{Defects, boundaries, and compactifications}

Prefactorization algebras apply to many different geometric situations,  so that it is easy to work on manifolds with boundaries or corners and to incorporate defects.
In this section we will briefly explain how other concepts from field theory are captured in the language of prefactorization algebras.

\subsection{Basic constructions}

In the one-dimensional case, it is straightforward to include a module for the algebra by including a boundary.

\begin{example}\label{ex:bdryasmodule}
Let $A$ be an associative algebra and $V$ a right $A$-module. 
This data determines a prefactorization algebra $\cF_{A,V}$ on the half-line $\RR_{\geq 0} = [0,\infty)$, as follows. To any open set $U$ in the interior $(0,\infty)$, we set $\cF_{A,V}(U) = A^{\rm fact}(U)$, and we use the multiplication map of $A$ to combine elements from disjoint intervals, as in section~\ref{sec:associativealgebras}. 
To an open set $U=[0,a)$, we set $\cF_{A,V}(U) = V$. 
The module structure determines the structure maps involving open sets that contain the boundary.
For instance, given open sets $U=[0,a)$ and $U'=(b,c)$, with $a < b$, and a larger open interval $U''=[0,d)$, with $c < d$, we have
\[
\cF_{A,V}(U) \otimes \cF_{A,V}(U') \cong V \otimes A \to V \cong \cF_{A,V}(U'').
\]
One can similarly work with a closed interval $[0,T]$ and describe incoming and outgoing states.
See \cite{ChiHohPin} for a detailed analysis of this situation using the BV formalism.~\hfill$\Diamond$
\end{example}

More generally, a QFT with a boundary condition gives rise to a factorization algebra on a manifold with boundary (in the case above, on $[0,\infty)$).
A similar construction can be used for a ``domain wall'' between two systems: for $t < 0$, one has an algebra $A_-$ and for $t > 0$, one has an algebra $A_+$, and for an open interval containing the origin, one has a bimodule.  

Maps between spaces let one produce new factorization algebras.

\begin{dfn}
Let $p: M \to N$ be a continuous map.
Let $\cF$ be a factorization algebra on $M$.
Its {\em pushforward} $p_*\cF$ is the factorization algebra on $N$ defined by
\[
p_*\cF(U) = \cF(p^{-1}U)
\]
on open sets and likewise for structure maps.
\end{dfn}

\begin{example}
Consider a prefactorization algebra $\cF$ for a theory on a manifold of the form $\RR \times N$, where we view $\RR$ as a time-like parameter and $N$ as space-like. Let us assume that $N$ is compat.
If we push forward along the projection $p: \RR \times N \to \RR$, we get a prefactorization algebra $p_* \cF$ on $\RR$ that can be seen as the observables of a quantum mechanical system.
In the case of a free scalar field, this push-forward operation gives the familiar Weyl algebra produced by canonical quantization.~\hfill$\Diamond$
\end{example}

Thus the pushforward also offers a language for {\it compactification}.  

This natural operation gives a view on the state-operator correspondence and on canonical quantization.

\begin{example}\label{ex:radquant}
We return to Euclidean field theory on $\RR^2$.  
Consider the map $p: \RR^2 \to [0,\infty)$ sending a point $(x,y)$ to its radius $r = \sqrt{x^2+y^2}$.
The interval $[0,a)$ has a disk centered at the origin as its preimage,
and any open interval $(a,b) \subset [0,\infty)$ has an annulus as its preimage.
The pushforward $p_* \Obs$ thus determines a prefactorization algebra whose value on an open interval $(a,b)$ behaves like an algebra of operators (those from radial quantization) and whose value on $[0,a)$ behaves like the state space of that algebra.
This state space is, however, the algebra of local operators in the two-dimensional theory.
In this sense we have a {\em state-operator} correspondence.~\hfill$\Diamond$
\end{example}

\subsection{An overview of related work}

In the topological setting, \cite{AFT2} and \cite{MW} have given a powerful formalism that characterizes the local algebraic structure of topological defects, 
generalizing the relationship between $E_n$ algebras and locally constant factorization algebras.
They explain, for instance, how to obtain knot invariants by this process.
This kind of relationship also offers a way to characterize defects using Koszul duality (see \cite{paquette2023koszul} for a nice discussion).

In \cite{CosYangian,CostelloStringMath}, Costello explained why line defects, in a gauge theory dubbed {\em 4-dimensional Chern-Simons theory}, satisfy the quantum Yang-Baxter equation and how the associated factorization algebra encodes the Yangian, a quantum group introduced by Drinfeld.
Subsequently, this framework has been expanded to encompass many further features of quantum integrable systems~\cite{CWY1,CWY2,CY, CosYag, CosGaiQ}.

In his thesis \cite{RabThesis}, Rabinovich proved an analog of Theorem~\ref{thm:main} for manifolds with boundary:
on a manifold with boundary, if a perturbative theory is topological in a direction normal to the boundary and a local boundary condition is imposed, then the observables form a factorization algebra whose observables in the bulk agree with those of \cite{CG2}.
This setting encompasses many examples (e.g., the Chern-Simons/WZW correspondence) but it also allows one to implement a standard ansatz for constructing defects (see \cite{CEG}, following \cite{KapDef} for the physical ansatz).
Rabinovich's approach should extend to situation where the theory is scale-equivariant (or conformal) in a direction normal to the boundary.

\section{Noether's theorem beyond the perturbative}
\label{sec:nonpert}

In this section we discuss how higher symmetries act on a QFT using {\em pre}\/factorization algebras (cf. \cite{Getal, Sharpe}),
inspired by our formulation of Noether currents using factorization algebras.
It can be seen as a conjectural formulation of a non-perturbative Noether theorem.  
The language here is a little more mathematically sophisticated, as it is important to use derived stacks.  
Our formulation should hold in either Euclidean or Lorentzian signature.   
Everything we discuss here has been formulated and proved for perturbative theories in~\cite{CG2}.  

From a mathematical perspective, there is an equivalence between an object with a symmetry and a family of objects.  
For example, let $G$ be a discrete group.  
To give a $G$-action on $\CC^n$ is then the same as giving a vector bundle $\mc{V}$ of rank $n$ on the classifying space $BG$, 
together with a framing of the fibre of $\mc{V}$ at the base point in~$BG$. 

More generally, given some derived stack $X$ with a base point $x \in X$, 
one can say that a vector space $V$ with {\em $X$-symmetry} is a vector bundle $\mc{V}$ on $X$ with an isomorphism~$\mc{V}_x \simeq V$. 

Similarly, for a quantum field theory, one should treat parameters in the theory (i.e., coupling constants) and symmetries on the same footing. 
We took this perspective in \cite{CG1,CG2};  
a somewhat similar perspective is taken in~\cite{Cordova:2019jnf}.  

Since a quantum field theory on a manifold $M$ is local on $M$, 
a symmetry should be given not just by a fixed space $X$ but by a sheaf of spaces $X$ on~$M$. 
Here are some examples to bear in mind.

\begin{example}
Let $\mc{L}$ be a sheaf of (homotopy) Lie algebras on a space $M$.  
There is a sheaf of formal derived stacks $B \mc{L}(U)$, the classifying space of the formal group associated to $\mc{L}(U)$.  	
A theory with $\mc{L}$-symmetry is the same as a family of theories over $B \mc{L}(U)$.  
This is the context relevant to Theorem~\ref{thm pert Noether}, our perturbative version of the Noether theorem.~\hfill$\Diamond$
\end{example}

\begin{example}
Let $\op{Conn}_G(U)$ be the stack of principal $G$-bundles on $U$ with a connection, 
up to gauge equivalence. 
As $U$ varies this construction forms a sheaf of stacks we call $\op{Conn}_G$.   
A theory $T$ that can be coupled to a background $G$-gauge field is the same as a family of theories $T^{\op{Conn}_G}$ over the sheaf $\op{Conn}_G$, which restricts to the original theory $T$ at the trivial connection.~\hfill$\Diamond$
\end{example}

\begin{rmk}
Physicists often conflate theories that have a $G$-symmetry and theories that live in a family over $\op{Conn}_G$. 
In general, these are not the same, although in physically relevant examples they tend to be.~\hfill$\Diamond$
\end{rmk}

We can now formulate a general notion.

\begin{dfn}
Let $X$ be a sheaf of (possibly infinite-dimensional) derived stacks over a manifold $M$, with a global section $x$.  
Let $\cT$ be a quantum field theory on $M$. 
Giving $X$-symmetry to $\cT$ means we give a family of theories $\cT^X$ over $X$ that restricts to $\cT$ at the section~$x$. 
\end{dfn}

\begin{rmk}
We will not attempt to be say here what infinite dimensional derived stack means in the setting of differential geometry, 
but see \cite{Steff, AlYo, Car} for steps in this direction.~\hfill$\Diamond$
\end{rmk}

In the language of factorization algebras of  \cite{CG2}, this definition means that the observables (or operators) on an open $U$ will be a kind of vector bundle over $X(U)$, 
and that the factorization product will be compatible with the sheaf structure on $X$.  
Since $X$ is a sheaf of stacks, this notion encompasses both varying the coupling constants as well as compatibility with symmetry.

One can formulate a rather general Noether ``theorem'' once one has this language.
More accurately, it is a proposal for capturing symmetries of the form just described,
and it is conjectural until an adequate framework for nonperturbative field theory emerges.
  
There are a few preliminary steps.  
We let $X_c(U)$ denote the space of sections of $X$ on $U$ that are equal to our base section $x$ outside a compact subset of $U$. 
It is the space of sections of $X$ with compact support.  

Note that if $U_1, U_2$ are disjoint open subsets of $V$, there is a natural map
\begin{equation} 
X_c(U_1) \times X_c(U_2) \to X_c(V)
\label{eqn:fact_space} 
\end{equation}
as follows.
If $s_1,s_2$ denote elements of $X_c(U_1)$, $X_c(U_2)$, respectively, 
then $s_i$ is equal to $x$ outside a compact subset $K_i$ of $U_i$. 
Hence $s_1$ extends to a section of $X$ on the open $U_2 \setminus K_2$ by setting it to be $x$ away from $K_1$.  
Similarly, $s_2$ extends to a section on the open $U_1 \setminus K_1$.  
These sections agree on their overlap and thus define a section on all of~$V$.
In short, we have shown that $X_c$ defines a \emph{prefactorization space}. 
 
\begin{example}
\label{ex:hfs}
Consider the case that the space-time manifold is $\RR$.
Let $G$ be a discrete group, 
and let $BG$ denote a classifying space with distinguished basepoint~$p$. 
Let $X$ be the sheaf of sections for the trivial bundle $\RR \times BG \to \RR$. 
For any open interval $I$, we find 
\[
X_c (I) = G, 
\]
because the space of compactly supported paths into $BG$ is equivalent to the based loop space $\Omega BG \simeq G$.
Alternatively, a compactly supported map into $BG$ determines a principal $G$-bundle on $I$ that is trivialized outside a compact set (i.e., the map equals $p$ outside that set), 
and hence the bundle is determined by its parallel transport from $0$ to~$1$.
Furthermore, if $I_1, I_2$ are disjoint intervals inside $I_3$, the factorization product
\[ 
X_c(I_1) \times X_c(I_2) \to X_c(I_3) 
\]
is the group multiplication $G \times G \to G$.

More generally, one can replace the real line with any manifold $M$, 
and this example shows how to encode {\em 0-form symmetries}.
If $N \subset M$ is a compact and connected submanifold of codimension one,
then a tubular neighborhood $U \supset N$ is diffeomorphic to $N \times \RR$.
Thus,
\[
\Map_c(U,BG) \cong \Map_c(N \times \RR,BG) \cong \Map(N, \Map_c(\RR,BG)) \simeq \Map(N, G),
\]
and as $G$ is discrete, we have $\Map(N,G) = G$.
In other words, we are labeling the codimension one submanifold $N$ by group elements.

By replacing $BG$ with $B^{k+1} A$, for $A$ an abelian group, 
one can similarly encode $k$-form symmetries on codimension $k+1$-submanifolds.~\hfill$\Diamond$
\end{example}

To build a prefactorization algebra, we need to apply some linearization functor that takes spaces to vector spaces. 
This functor must be covariant.  
A natural functor to use is \emph{distributions}, the (correct) linear dual to functions;    
since functions are contravariant, distributions are covariant. 
(In the setting of derived differential geometry, a careful treatment of distributions is not yet available, but one should expect it to exist on structural grounds.)

Let $\Dist(X_c(U))$ denote the space of distributions on $X_c(U)$, and let $\Dist_{X_c}$ denote the prefactorization algebra that assigns $\Dist(X_c(U))$ to each open set~$U$.

\begin{example}[Example \ref{ex:hfs} continued]
For $G$ discrete, we care about $BG$ as a homotopy type, and likewise for $X_c(U)$.
The distributions on a homotopy type are encoded by the singular chain complex $C_*(X_c(U))$.
(Functions are encoded by the singular cochains.)
When the spacetime is the real line $\RR$, we find that 
\[
\Dist(\cB\cG_c)(I) \cong C_*(G) \simeq \CC G, 
\]
so the we recover the {\em group ring} as the linearized symmetries.~\hfill$\Diamond$
\end{example}

Roughly speaking, the Noether theorem should say that when a theory has $X$-symmetry, 
the prefactorization algebra of observables receives a map from $\Dist_{X_c}$.  
There is one subtlety, though: we need to take into account 't Hooft anomalies, 
which in a perturbative treatment lead to a central extension to the algebra.  

To understand the 't Hooft anomaly, let us return to case that the space-time manifold is $\RR$ and $X$ is the constant sheaf $BG$, the classifying space of a discrete group. 
Here, we expect the 't Hooft anomaly to yield a central extension of $G$, 
namely a group $\widehat{G}$ living in an exact sequence
\[ 
	1 \to \CC^\times \to \widehat{G} \to G \to 1. 
\]
The central extension $\widehat{G}$ is a principal $\CC^\times$-bundle over $G$, i.e., a line bundle $\mc{L}$. 
The fact that $\widehat{G}$ is itself a group is reflected in the fact that the line bundle $\mc{L}$ is {\em multiplicative}: 
under the group multiplication map $m : G \times G \to G$,
there is an isomorphism
\[ 
	m^\ast \mc{L} \simeq \mc{L} \boxtimes \mc{L}. 
\]
(This isomorphism must be associative in the natural sense).
Since the product on $G$ comes from the prefactorization structure on $X_c$, 
we introduce the following definition in general. 

\begin{dfn}
Given the prefactorization space $X_c$ as above, a {\em factorizing line bundle} is a line bundle $\mc{L}(U)$ on every $X_c(U)$, with coherent isomorphisms 
\[ 
m^\ast \mc{L}(V) \simeq \mc{L}(U_1) \boxtimes \mc{L}(U_2) 
\]
associated to the factorization products~\eqref{eqn:fact_space}. 
\end{dfn}

Equivalently, we note that sending every $U$ to $B \CC^\times$ defines a prefactorization space on $M$, where the factorization product is defined using the fact that $B \CC^\times$ is an abelian group in stacks.   A factorizing line bundle is a map of prefactorization spaces from $X_c$ to~$B \CC^\times$.

Finally, we can formulate the non-perturbative analog of the Noether theorem:

\begin{conjecture*}
Let $\cT$ be a quantum field theory on a manifold $M$, and let $X$ be a sheaf of derived stacks on~$M$.  
If $\cT$ has $X$-symmetry, then there exists a factorizing line bundle $\mc{L}$ on the prefactorization space $X_c$ and a map of prefactorization algebras
\[
J: \Dist_{X_c}^{\mc{L}} \to \Obs_\cT. 
\]
from distributions on $X$, twisted by $\mc{L}$, to $\Obs_\cT$.	
\end{conjecture*}

This relationship should hold in both Lorentzian and Euclidean signature.

In the case that $\cT$ is a perturbative theory and we work in Euclidean signature, and $X$ is a sheaf of formal derived stacks (completed along the section $x$), then this conjecture is the Noether theorem proved in~\cite{CG2}.  

\begin{rmk}
In Lorentzian signature, it is important to note that $\Dist_{X_c}^{\mc{L}}$ does \emph{not} normally satisfy the Cauchy constancy axiom of \cite{Benini_2019}. 
This axiom fails even in the basic case when $X = \op{Conn}_G$ is the stack of $G$-connections,
roughly because a $G$-connection does not satisfy any PDE whatsoever and so is not determined by initial data.  
Thus, this Noether construction cannot be entirely formulated in AQFT in its standard form.~\hfill$\Diamond$
\end{rmk}

Let us give some examples of these nonperturbative Noether currents.

\begin{example}
Consider a quantum mechanical system with $G$-symmetry, where $G$ is a discrete group.  
As we have seen, for any interval, $\op{Conn}_{G,c}(I) \simeq G$.  
The non-perturbative Noether theorem tells us that there is central extension of $G$ and a ring homomorphism from the twisted group algebra $\CC_{L} [G]$ to the algebra of operators of the system.

In the case $G$ is a Lie group, a similar statement applies, although the twisted group algebra should be interpreted as the algebra of distributions on $G$ under convolution.  
Distributions supported at the identity in $G$ give the (twisted) universal enveloping algebra of $\mathfrak{g}$, recovering the perturbative statement of Noether's theorem.~\hfill$\Diamond$
\end{example}

\begin{example}
Let $T = (\CC^\times)^n$ be an Abelian complex algebraic group.  Suppose that $T$ acts as symmetries on a chiral conformal field theory on a Riemann surface $M$. For $U \subset M$, let $\Bun_T(U)$ be the stack of holomorphic $T$-bundles on $U$.  This has a basepoint given by the trivial bundle.

Let $\mc{F}$ be the factorization algebra associated to a chiral CFT with $T$-symmetry. Such a CFT can be coupled to a background holomorphic $T$-bundle and so has $\Bun_T$-symmetry.

Now observe that for $U$ a very small disk --- in fact, a formal disk $\widehat{\DD}$ --- the space $\Bun_{T,c}(\widehat{\DD})$ is the affine (or loop) Grassmannian ${\rm Gr}_T$. 
It has components labelled by~$\pi_1(T) \cong \ZZ^n$, a lattice.
\cite{BD} described, in their algebraic context, this factorization space, now known as the Beilinson-Drinfeld Grassmannian.

Noether's theorem then says there will be a map from $\Dist(\Bun_{T,c},L)$ to observable of the chiral CFT, for a line bundle on $L$.  
There is a factorizing line bundle on $\Bun_{T,c}$ associated to a pairing on the lattice~$\pi_1(T)$.
Beilinson and Drinfeld have shown that $\Dist(\Bun_{G,c}, L)$ encodes the lattice vertex algebra associated to the lattice~$\pi_1(T)$.   
	
The map $J: {\rm Dist}(\Bun_{T,c}, L) \to \mc{F}$ then encodes how this WZW model acts on the chiral CFT.
Note that in this example, the perturbative Noether theorem is much weaker: 
it only yields a map from the Heisenberg subalgebra of the lattice vertex algebra.~\hfill$\Diamond$
\end{example}

\begin{example}
An important special case of the previous example is given when $T = \CC^\times$, and $\mc{F}$ is the factorization algebra associated to a pair  of chiral free fermion fields  $\psi, \psi'$ with Lagrangian $\int \psi \bar{\partial} \psi'$. (Physicists would call this system a chiral complex fermion.)   
In this case, the map
\[ 
J : \Dist (\Bun_{T,c}, L) \to \mc{F}
\]
is an isomorphism, and it encodes  the {\em boson-fermion correspondence}.  
It is essential that we use the non-perturbative Noether theorem for this statement, 
as the fermion fields $\psi$ and $\psi'$ come from components of the affine Grassmanian disjoint from the identity component.~\hfill$\Diamond$
\end{example}

\begin{example}
Consider any quantum field theory on a manifold $M$ of dimension $n$ that can be coupled to a background $U(1)$ gauge field, i.e., has $\op{Conn}_{U(1)}$-symmetry. 
For an open subset $U \subset M$, the space $\op{Conn}_{U(1),c} (U)$
has components labelled by $H^2_c(U)$, which, by Poincar\'e duality, is isomorphic to $H_{n-2}(U)$.  
The non-perturbative Noether theorem thus supplements the ordinary local Noether currents by non-local defects living on codimension $2$ subspaces.
This observation is behind several of the higher-form symmetries discussed in~\cite{Getal}.~\hfill$\Diamond$
\end{example}

There are further examples of the non-perturbative Noether's theorem associated to homotopy types, 
as we discussed in Example~\ref{ex:hfs} and which we can quickly generalize now.

Let $(X,x)$ be a pointed topological space. 
The prefactorization space $X_c$  on a manifold $M$ assigns compactly supported maps into $X$; 
the natural linearization is given by singular chains, giving a prefactorization algebra~$C_\ast(X_c)$. 
This prefactorization algebra $\op{Chains}(\cX_c) = C_*(\Map_c(U,X))$ with values in chain complexes is the {\em higher $X$-symmetry algebra}.  
(It is already well-understood in topology via {\em nonabelian Poincar\'e duality} of \cite{SegFA, McDuff, Salv, LurieHA, AF1}.)
Symmetries associated to such homotopy types encode generalized global symmetries. 
For $X = BG$,  as we have already discussed, one encodes 0-form symmetry by a Noether map~$J$.

\begin{example}
Let $G$ be a compact semisimple Lie group $G$, and let $Z$ be the center of $G$.
Using $X = B^2 Z$, one can encode the {\em center symmetry} of a $G$-gauge theory by such a map.  
Indeed, in this case, the relevant prefactorization space has the feature that 
\[ 
\pi_0(\Map_c(U, B^2 Z)) = H^2_c(U,Z) = H_{n-2}(U,Z). 
\]
We thus find symmetries associated to neighborhoods of codimension $2$ manifolds. 

Any gauge theory with $G$-symmetry can be coupled to a background field, which is a map to $B^2 Z$. It is perhaps easiest to see this using the \v{C}ech picture: 
given a \v{C}ech $2$-cocycle valued in $Z$, one can modify the notion of principal $G$-bundle with connection by asking that the failure of the  transition functions to satisfy the cocycle condition is given by the chosen $Z$-valued $2$-cocycle. 
Taking such  twisted principal $G$-bundles with connection as fields modifies the gauge theory. 

The Noether theorem then gives us a homomorphism from the prefactorization algebra $C_*(\op{Maps}_c(B^2 Z))$ to observables of the theory.  These observables are the standard higher symmetries associated to the center of the group, as explained in~\cite{Getal}.~\hfill$\Diamond$
\end{example}

There is a great deal of further work to be done to flesh out this conjectural Noether theorem.  
In particular, we have been very vague about what one means by distributions on the infinite-dimensional derived stack that appear in the formulation of the conjecture, 
and we have explored few of the symmetry prefactorization algebras that appear in the theorem.  

\bibliographystyle{agsm}
\bibliography{Proposal}

@article {CWY2,
    AUTHOR = {Costello, Kevin and Witten, Edward and Yamazaki, Masahito},
     TITLE = {Gauge theory and integrability, {II}},
   JOURNAL = {ICCM Not.},
  FJOURNAL = {ICCM Notices. Notices of the International Congress of Chinese
              Mathematicians},
    VOLUME = {6},
      YEAR = {2018},
    NUMBER = {1},
     PAGES = {120--146},
      ISSN = {2326-4810},
   MRCLASS = {81T13 (16T25 53C05 81T45 82B23)},
  MRNUMBER = {3855890},
MRREVIEWER = {Derek G. Harland},
       DOI = {10.4310/ICCM.2018.v6.n1.a7},
       URL = {https://doi.org/10.4310/ICCM.2018.v6.n1.a7},
}

@unpublished {CY,
    AUTHOR = {Costello, Kevin and Yamazaki, Masahito},
     TITLE = {Gauge theory and integrability, {III}},
NOTE = {Available at \url{https://arxiv.org/abs/1908.02289}},
}

@article {MW,
    AUTHOR = {Morrison, Scott and Walker, Kevin},
     TITLE = {Blob homology},
   JOURNAL = {Geom. Topol.},
  FJOURNAL = {Geometry \& Topology},
    VOLUME = {16},
      YEAR = {2012},
    NUMBER = {3},
     PAGES = {1481--1607},
      ISSN = {1465-3060},
   MRCLASS = {57R56 (18G35)},
  MRNUMBER = {2978449},
MRREVIEWER = {Yasuyuki Kawahigashi},
       DOI = {10.2140/gt.2012.16.1481},
       URL = {http://dx.doi.org/10.2140/gt.2012.16.1481},
}

@inproceedings {CWY1,
    AUTHOR = {Costello, Kevin and Witten, Edward and Yamazaki, Masahito},
     TITLE = {Gauge theory and integrability, {I}},
 BOOKTITLE = {Proceedings of the {S}eventh {I}nternational {C}ongress of
              {C}hinese {M}athematicians, {V}ol. {I}},
    SERIES = {Adv. Lect. Math. (ALM)},
    VOLUME = {43},
     PAGES = {17--151},
 PUBLISHER = {Int. Press, Somerville, MA},
      YEAR = {2019},
   MRCLASS = {81R12 (70S15)},
  MRNUMBER = {3971869},
}

@article {LiVOA,
    AUTHOR = {Li, Si},
     TITLE = {Vertex algebras and quantum master equation},
   JOURNAL = {J. Differential Geom.},
  FJOURNAL = {Journal of Differential Geometry},
    VOLUME = {123},
      YEAR = {2023},
    NUMBER = {3},
     PAGES = {461--521},
      ISSN = {0022-040X,1945-743X},
   MRCLASS = {53D50 (17B69 81R15 81T40)},
  MRNUMBER = {4584859},
       DOI = {10.4310/jdg/1683307007},
       URL = {https://doi.org/10.4310/jdg/1683307007},
}

@book {CosBook,
    AUTHOR = {Costello, Kevin},
     TITLE = {Renormalization and effective field theory},
    SERIES = {Mathematical Surveys and Monographs},
    VOLUME = {170},
 PUBLISHER = {American Mathematical Society},
   ADDRESS = {Providence, RI},
      YEAR = {2011},
     PAGES = {viii+251},
      ISBN = {978-0-8218-5288-0},
   MRCLASS = {81T15 (58Jxx 81T70)},
  MRNUMBER = {2778558},
}

@article {Cordova:2019jnf,
    AUTHOR = {C\'{o}rdova, Clay and Freed, Daniel S. and Lam, Ho Tat and
              Seiberg, Nathan},
     TITLE = {Anomalies in the space of coupling constants and their
              dynamical applications {I}},
   JOURNAL = {SciPost Phys.},
  FJOURNAL = {SciPost Physics},
    VOLUME = {8},
      YEAR = {2020},
    NUMBER = {1},
     PAGES = {Paper No. 001, 57},
      ISSN = {2542-4653},
   MRCLASS = {81T50 (81T70)},
  MRNUMBER = {4158234},
MRREVIEWER = {Dmitri\ V.\ Vassilevich},
       DOI = {10.21468/SciPostPhys.8.1.001},
       URL = {https://doi.org/10.21468/SciPostPhys.8.1.001},
}

@book {CG1,
    AUTHOR = {Costello, Kevin and Gwilliam, Owen},
     TITLE = {Factorization algebras in quantum field theory. {V}ol. {1}},
    SERIES = {New Mathematical Monographs},
    VOLUME = {31},
 PUBLISHER = {Cambridge University Press, Cambridge},
      YEAR = {2017},
     PAGES = {ix+387},
      ISBN = {978-1-107-16310-2},
   MRCLASS = {81-01 (17B69 18D50 53D55 81R05 81R10)},
  MRNUMBER = {3586504},
       DOI = {10.1017/9781316678626},
       URL = {http://dx.doi.org/10.1017/9781316678626},
}

@book {CG2,
    AUTHOR = {Costello, Kevin and Gwilliam, Owen},
     TITLE = {Factorization algebras in quantum field theory. {V}ol. 2},
    SERIES = {New Mathematical Monographs},
    VOLUME = {41},
 PUBLISHER = {Cambridge University Press, Cambridge},
      YEAR = {2021},
     PAGES = {xiii+402},
      ISBN = {978-1-107-16315-7; 978-1-009-00616-3},
   MRCLASS = {81-02 (18M99 70S10 81R10 81T70)},
  MRNUMBER = {4300181},
MRREVIEWER = {Domenico\ Fiorenza},
       DOI = {10.1017/9781316678664},
       URL = {https://doi.org/10.1017/9781316678664},
}

@article{GLL,
	AUTHOR = {Grady, Ryan E. and Li, Qin and Li, Si},
     TITLE = {Batalin-{V}ilkovisky quantization and the algebraic index},
   JOURNAL = {Adv. Math.},
  FJOURNAL = {Advances in Mathematics},
    VOLUME = {317},
      YEAR = {2017},
     PAGES = {575--639},
      ISSN = {0001-8708},
   MRCLASS = {53D55 (58J20 81Q30 81T15)},
  MRNUMBER = {3682678},
MRREVIEWER = {Andrea D'Agnolo},
       DOI = {10.1016/j.aim.2017.07.007},
       URL = {https://doi.org/10.1016/j.aim.2017.07.007},
}

@article {ChanLeungLi,
    AUTHOR = {Chan, Kwokwai and Leung, Naichung Conan and Li, Qin},
     TITLE = {B{V} quantization of the {R}ozansky-{W}itten model},
   JOURNAL = {Comm. Math. Phys.},
  FJOURNAL = {Communications in Mathematical Physics},
    VOLUME = {355},
      YEAR = {2017},
    NUMBER = {1},
     PAGES = {97--144},
      ISSN = {0010-3616,1432-0916},
   MRCLASS = {53D50 (81T20)},
  MRNUMBER = {3670730},
MRREVIEWER = {Yisong\ Yang},
       DOI = {10.1007/s00220-017-2924-8},
       URL = {https://doi.org/10.1007/s00220-017-2924-8},
}

@article {AF1,
    AUTHOR = {Ayala, David and Francis, John},
     TITLE = {Factorization homology of topological manifolds},
   JOURNAL = {J. Topol.},
  FJOURNAL = {Journal of Topology},
    VOLUME = {8},
      YEAR = {2015},
    NUMBER = {4},
     PAGES = {1045--1084},
      ISSN = {1753-8416},
   MRCLASS = {57R56 (55U40 57N35)},
  MRNUMBER = {3431668},
       DOI = {10.1112/jtopol/jtv028},
       URL = {http://dx.doi.org/10.1112/jtopol/jtv028},
}

@unpublished{AFCob,
AUTHOR = {Ayala, David and Francis, John},
TITLE = {The cobordism hypothesis},
NOTE = {Available at \url{https://arxiv.org/abs/1705.02240}},
}

@unpublished {LurieHA,
    AUTHOR = {Lurie, Jacob},
     TITLE = {Higher {A}lgebra},
NOTE = {Available at \url{https://www.math.ias.edu/~lurie/}},
}

@article {LiLi,
    AUTHOR = {Li, Qin and Li, Si},
     TITLE = {On the {B}-twisted topological sigma model and {C}alabi-{Y}au
              geometry},
   JOURNAL = {J. Differential Geom.},
  FJOURNAL = {Journal of Differential Geometry},
    VOLUME = {102},
      YEAR = {2016},
    NUMBER = {3},
     PAGES = {409--484},
      ISSN = {0022-040X},
   MRCLASS = {53D50 (53D37)},
  MRNUMBER = {3466804},
       URL = {http://projecteuclid.org/euclid.jdg/1456754015},
}

@article{GGW,
AUTHOR = {Gwilliam, Owen and Gorbounov, Vassily and Williams, Brian},
TITLE = {{C}hiral differential operators via quantization of the holomorphic $\sigma$-model},
JOURNAL = {Ast\'{e}risque},
  FJOURNAL = {Ast\'{e}risque},
    VOLUME = {419},
      YEAR = {2020},
}

@article {AFT2,
    AUTHOR = {Ayala, David and Francis, John and Tanaka, Hiro Lee},
     TITLE = {Factorization homology of stratified spaces},
   JOURNAL = {Selecta Math. (N.S.)},
  FJOURNAL = {Selecta Mathematica. New Series},
    VOLUME = {23},
      YEAR = {2017},
    NUMBER = {1},
     PAGES = {293--362},
      ISSN = {1022-1824},
   MRCLASS = {57P05 (55N40 57N80 57R40 57R56)},
  MRNUMBER = {3595895},
       DOI = {10.1007/s00029-016-0242-1},
       URL = {http://dx.doi.org/10.1007/s00029-016-0242-1},
}

@unpublished {CosYangian,
   AUTHOR = {Costello, Kevin},
     TITLE = {{S}upersymmetric gauge theory and the {Y}angian},
 NOTE = {Available at \url{http://arxiv.org/abs/1303.2632}},
}

@book {BD,
    AUTHOR = {Beilinson, Alexander and Drinfeld, Vladimir},
     TITLE = {Chiral algebras},
    SERIES = {American Mathematical Society Colloquium Publications},
    VOLUME = {51},
 PUBLISHER = {American Mathematical Society, Providence, RI},
      YEAR = {2004},
     PAGES = {vi+375},
      ISBN = {0-8218-3528-9},
   MRCLASS = {17Bxx (14F43)},
  MRNUMBER = {2058353},
MRREVIEWER = {Francisco J. Plaza Mart\'\i n},
       DOI = {10.1090/coll/051},
       URL = {http://dx.doi.org/10.1090/coll/051},
}

@incollection {SegalCFT,
    AUTHOR = {Segal, G. B.},
     TITLE = {The definition of conformal field theory},
 BOOKTITLE = {Differential geometrical methods in theoretical physics
              ({C}omo, 1987)},
    SERIES = {NATO Adv. Sci. Inst. Ser. C Math. Phys. Sci.},
    VOLUME = {250},
     PAGES = {165--171},
 PUBLISHER = {Kluwer Acad. Publ., Dordrecht},
      YEAR = {1988},
   MRCLASS = {58D30 (14H15 32G15 81E05 81E40)},
  MRNUMBER = {981378},
MRREVIEWER = {Yukihiko Namikawa},
}

@article {RabAxial,
    AUTHOR = {Rabinovich, Eugene},
     TITLE = {A mathematical analysis of the axial anomaly},
   JOURNAL = {Lett. Math. Phys.},
  FJOURNAL = {Letters in Mathematical Physics},
    VOLUME = {109},
      YEAR = {2019},
    NUMBER = {5},
     PAGES = {1055--1117},
      ISSN = {0377-9017},
   MRCLASS = {81T15 (18G40 35K08 35Q40 81T18 81T20 81T50)},
  MRNUMBER = {3946486},
       DOI = {10.1007/s11005-018-1142-4},
       URL = {https://doi.org/10.1007/s11005-018-1142-4},
}

@article {HK,
    AUTHOR = {Hennion, Benjamin and Kapranov, Mikhail},
     TITLE = {Gelfand-{F}uchs cohomology in algebraic geometry and
              factorization algebras},
   JOURNAL = {J. Amer. Math. Soc.},
  FJOURNAL = {Journal of the American Mathematical Society},
    VOLUME = {36},
      YEAR = {2023},
    NUMBER = {2},
     PAGES = {311--396},
      ISSN = {0894-0347,1088-6834},
   MRCLASS = {14F10 (17B56)},
  MRNUMBER = {4536901},
MRREVIEWER = {Dingxin\ Zhang},
       DOI = {10.1090/jams/1001},
       URL = {https://doi.org/10.1090/jams/1001},
}

@Article{gaiotto,
author="Bullimore, Mathew
and Dimofte, Tudor
and Gaiotto, Davide
and Hilburn, Justin",
title="Boundaries, mirror symmetry, and symplectic duality in 3d {N}=4 gauge theory",
journal="Journal of High Energy Physics",
year="2016",
month="Oct",
day="20",
volume="2016",
number="10",
pages="108",
issn="1029-8479",
doi="10.1007/JHEP10(2016)108",
url="https://doi.org/10.1007/JHEP10(2016)108"
}

@article{ES,
AUTHOR = {Elliott, Chris and Safronov, Pavel},
     TITLE = {Topological twists of supersymmetric algebras of observables},
   JOURNAL = {Comm. Math. Phys.},
  FJOURNAL = {Communications in Mathematical Physics},
    VOLUME = {371},
      YEAR = {2019},
    NUMBER = {2},
     PAGES = {727--786},
      ISSN = {0010-3616},
   MRCLASS = {81T60 (17B37 18D50 18F20)},
  MRNUMBER = {4019918},
       DOI = {10.1007/s00220-019-03393-9},
       URL = {https://doi.org/10.1007/s00220-019-03393-9},
}

@phdthesis{Scheim,
  title={Factorization homology as a fully extended topological field theory},
  author={Scheimbauer, Claudia},
  year={2014},
  school={ETH Zurich}
}

@article {RabDet,
    AUTHOR = {Rabinovich, Eugene},
     TITLE = {The {B}atalin-{V}ilkovisky formalism and the determinant line
              bundle},
   JOURNAL = {J. Geom. Phys.},
  FJOURNAL = {Journal of Geometry and Physics},
    VOLUME = {156},
      YEAR = {2020},
%     PAGES = {103792, 18},
      ISSN = {0393-0440},
   MRCLASS = {81T70 (58J52 81T50)},
  MRNUMBER = {4118887},
       DOI = {10.1016/j.geomphys.2020.103792},
       URL = {https://doi.org/10.1016/j.geomphys.2020.103792},
}

@incollection {LurieTFT,
    AUTHOR = {Lurie, Jacob},
     TITLE = {On the classification of topological field theories},
 BOOKTITLE = {Current developments in mathematics, 2008},
     PAGES = {129--280},
 PUBLISHER = {Int. Press, Somerville, MA},
      YEAR = {2009},
   MRCLASS = {57R56 (18D10 18G30 57R15 57R75)},
  MRNUMBER = {2555928 (2010k:57064)},
MRREVIEWER = {Julia Bergner},
}

@article {BBJ1,
    AUTHOR = {Ben-Zvi, David and Brochier, Adrien and Jordan, David},
     TITLE = {Integrating quantum groups over surfaces},
   JOURNAL = {J. Topol.},
  FJOURNAL = {Journal of Topology},
    VOLUME = {11},
      YEAR = {2018},
    NUMBER = {4},
     PAGES = {874--917},
      ISSN = {1753-8416},
   MRCLASS = {14D24 (14D23 16T25 18D10 57R56 57T05)},
  MRNUMBER = {3847209},
MRREVIEWER = {Jason Stuart Hanson},
       DOI = {10.1112/topo.12072},
       URL = {https://doi.org/10.1112/topo.12072},
}

@book {Li,
    AUTHOR = {Li, Si},
     TITLE = {Calabi-{Y}au {G}eometry and {H}igher {G}enus {M}irror
              {S}ymmetry},
      NOTE = {Thesis (Ph.D.)--Harvard University},
 PUBLISHER = {ProQuest LLC, Ann Arbor, MI},
      YEAR = {2011},
     PAGES = {174},
      ISBN = {978-1124-73644-0},
   MRCLASS = {Thesis},
  MRNUMBER = {2898602},
       URL =
              {http://gateway.proquest.com/openurl?url_ver=Z39.88-2004&rft_val_fmt=info:ofi/fmt:kev:mtx:dissertation&res_dat=xri:pqdiss&rft_dat=xri:pqdiss:3462673},
}

@article {AFPK,
    AUTHOR = {Ayala, David and Francis, John},
     TITLE = {Poincar\'{e}/{K}oszul duality},
   JOURNAL = {Comm. Math. Phys.},
  FJOURNAL = {Communications in Mathematical Physics},
    VOLUME = {365},
      YEAR = {2019},
    NUMBER = {3},
     PAGES = {847--933},
      ISSN = {0010-3616,1432-0916},
   MRCLASS = {55U40 (17B70 55P65 57R56)},
  MRNUMBER = {3916983},
MRREVIEWER = {Martin\ Frankland},
       DOI = {10.1007/s00220-019-03311-z},
       URL = {https://doi.org/10.1007/s00220-019-03311-z},
}

@article {AxeSing,
    AUTHOR = {Axelrod, Scott and Singer, I. M.},
     TITLE = {Chern-{S}imons perturbation theory. {II}},
   JOURNAL = {J. Differential Geom.},
  FJOURNAL = {Journal of Differential Geometry},
    VOLUME = {39},
      YEAR = {1994},
    NUMBER = {1},
     PAGES = {173--213},
      ISSN = {0022-040X},
     CODEN = {JDGEAS},
   MRCLASS = {58G26 (57R20 81T18 81T60)},
  MRNUMBER = {1258919},
MRREVIEWER = {Steven Rosenberg},
       URL = {http://projecteuclid.org/euclid.jdg/1214454681},
}

@incollection {KonECM,
    AUTHOR = {Kontsevich, Maxim},
     TITLE = {Feynman diagrams and low-dimensional topology},
 BOOKTITLE = {First {E}uropean {C}ongress of {M}athematics, {V}ol.\ {II}
              ({P}aris, 1992)},
    SERIES = {Progr. Math.},
    VOLUME = {120},
     PAGES = {97--121},
 PUBLISHER = {Birkh\"auser, Basel},
      YEAR = {1994},
   MRCLASS = {57R57 (14H15 32G15 57M25)},
  MRNUMBER = {1341841},
MRREVIEWER = {Anatoly Libgober},
}

@article {GwiRej1,
    AUTHOR = {Gwilliam, Owen and Rejzner, Kasia},
     TITLE = {Relating nets and factorization algebras of observables: free
              field theories},
   JOURNAL = {Comm. Math. Phys.},
  FJOURNAL = {Communications in Mathematical Physics},
    VOLUME = {373},
      YEAR = {2020},
    NUMBER = {1},
     PAGES = {107--174},
      ISSN = {0010-3616},
   MRCLASS = {81T05 (18M15 81T15)},
  MRNUMBER = {4050093},
       DOI = {10.1007/s00220-019-03652-9},
       URL = {https://doi.org/10.1007/s00220-019-03652-9},
}

@article {Knudsen,
    AUTHOR = {Knudsen, Ben},
     TITLE = {Betti numbers and stability for configuration spaces via
              factorization homology},
   JOURNAL = {Algebr. Geom. Topol.},
  FJOURNAL = {Algebraic \& Geometric Topology},
    VOLUME = {17},
      YEAR = {2017},
    NUMBER = {5},
     PAGES = {3137--3187},
      ISSN = {1472-2747},
   MRCLASS = {57R19 (17B56 55R80)},
  MRNUMBER = {3704255},
MRREVIEWER = {Daniel C. Cohen},
       DOI = {10.2140/agt.2017.17.3137},
       URL = {https://doi.org/10.2140/agt.2017.17.3137},
}

@incollection {AFprim,
    AUTHOR = {Ayala, David and Francis, John},
     TITLE = {A factorization homology primer},
 BOOKTITLE = {Handbook of homotopy theory},
    SERIES = {CRC Press/Chapman Hall Handb. Math. Ser.},
     PAGES = {39--101},
 PUBLISHER = {CRC Press, Boca Raton, FL},
      YEAR = {2020},
      ISBN = {978-0-815-36970-7},
   MRCLASS = {55N40 (57N35 57R56)},
  MRNUMBER = {4197982},
}

@article {Benini_2019,
    AUTHOR = {Benini, Marco and Perin, Marco and Schenkel, Alexander},
     TITLE = {Model-independent comparison between factorization algebras
              and algebraic quantum field theory on {L}orentzian manifolds},
   JOURNAL = {Comm. Math. Phys.},
  FJOURNAL = {Communications in Mathematical Physics},
    VOLUME = {377},
      YEAR = {2020},
    NUMBER = {2},
     PAGES = {971--997},
      ISSN = {0010-3616,1432-0916},
   MRCLASS = {81T05 (16W10 18M20 53C50 81R15)},
  MRNUMBER = {4115011},
MRREVIEWER = {Yoh\ Tanimoto},
       DOI = {10.1007/s00220-019-03561-x},
       URL = {https://doi.org/10.1007/s00220-019-03561-x},
}

@article {BBJ2,
    AUTHOR = {Ben-Zvi, David and Brochier, Adrien and Jordan, David},
     TITLE = {Quantum character varieties and braided module categories},
   JOURNAL = {Selecta Math. (N.S.)},
  FJOURNAL = {Selecta Mathematica. New Series},
    VOLUME = {24},
      YEAR = {2018},
    NUMBER = {5},
     PAGES = {4711--4748},
      ISSN = {1022-1824,1420-9020},
   MRCLASS = {17B37 (16T05 18D10)},
  MRNUMBER = {3874702},
MRREVIEWER = {Kevin\ D.\ Coulembier},
       DOI = {10.1007/s00029-018-0426-y},
       URL = {https://doi.org/10.1007/s00029-018-0426-y},
}

@misc{brochier2023classification,
      title={A Classification of Modular Functors via Factorization Homology}, 
      author={Adrien Brochier and Lukas Woike},
      year={2023},
      eprint={2212.11259},
      archivePrefix={arXiv},
      primaryClass={math.QA}
}

@article {Brochier_2021,
    AUTHOR = {Brochier, Adrien and Jordan, David and Snyder, Noah},
     TITLE = {On dualizability of braided tensor categories},
   JOURNAL = {Compos. Math.},
  FJOURNAL = {Compositio Mathematica},
    VOLUME = {157},
      YEAR = {2021},
    NUMBER = {3},
     PAGES = {435--483},
      ISSN = {0010-437X,1570-5846},
   MRCLASS = {17B37 (16D90 18M15 57K16 57K31)},
  MRNUMBER = {4228258},
MRREVIEWER = {Philsang\ Yoo},
       DOI = {10.1112/s0010437x20007630},
       URL = {https://doi.org/10.1112/s0010437x20007630},
}

@article {cooke2019excision,
    AUTHOR = {Cooke, Juliet},
     TITLE = {Excision of skein categories and factorisation homology},
   JOURNAL = {Adv. Math.},
  FJOURNAL = {Advances in Mathematics},
    VOLUME = {414},
      YEAR = {2023},
     PAGES = {Paper No. 108848, 51},
      ISSN = {0001-8708,1090-2082},
   MRCLASS = {57K18 (18M30 55N40)},
  MRNUMBER = {4536120},
MRREVIEWER = {Inbar\ Klang},
       DOI = {10.1016/j.aim.2022.108848},
       URL = {https://doi.org/10.1016/j.aim.2022.108848},
}

@article {GuiLi1,
    AUTHOR = {Gui, Zhengping and Li, Si and Xu, Kai},
     TITLE = {Geometry of localized effective theories, exact semi-classical
              approximation and the algebraic index},
   JOURNAL = {Comm. Math. Phys.},
  FJOURNAL = {Communications in Mathematical Physics},
    VOLUME = {382},
      YEAR = {2021},
    NUMBER = {1},
     PAGES = {441--483},
      ISSN = {0010-3616,1432-0916},
   MRCLASS = {81Q20 (17B81 58J20 81S40 81T45)},
  MRNUMBER = {4223479},
MRREVIEWER = {Stanislav\ Z.\ Pakuliak},
       DOI = {10.1007/s00220-021-03944-z},
       URL = {https://doi.org/10.1007/s00220-021-03944-z},
}

@article {keller2023finite,
    AUTHOR = {Keller, Corina and M\"{u}ller, Lukas},
     TITLE = {Finite symmetries of quantum character stacks},
   JOURNAL = {Theory Appl. Categ.},
  FJOURNAL = {Theory and Applications of Categories},
    VOLUME = {39},
      YEAR = {2023},
     PAGES = {Paper No. 3, 51--97},
      ISSN = {1201-561X},
   MRCLASS = {57K16 (18M15 18M60)},
  MRNUMBER = {4542026},
}

@article {paquette2023koszul,
    AUTHOR = {Paquette, Natalie M. and Williams, Brian R.},
     TITLE = {Koszul duality in quantum field theory},
   JOURNAL = {Confluentes Math.},
  FJOURNAL = {Confluentes Mathematici},
    VOLUME = {14},
      YEAR = {2022},
    NUMBER = {2},
     PAGES = {87--138},
      ISSN = {1793-7434},
   MRCLASS = {81T35 (18M70 81T13 81T30 81T60)},
  MRNUMBER = {4561878},
}

@book {Tetal,
    AUTHOR = {Tanaka, Hiro Lee},
     TITLE = {Lectures on factorization homology, {$\infty$}-categories, and
              topological field theories},
    SERIES = {SpringerBriefs in Mathematical Physics},
    VOLUME = {39},
      NOTE = {With contributions by Araminta Amabel, Artem Kalmykov and
              Lukas M\"{u}ller},
 PUBLISHER = {Springer, Cham},
      YEAR = {2020},
     PAGES = {xii+84},
      ISBN = {978-3-030-61163-7; 978-3-030-61162-0},
   MRCLASS = {55U35 (18M20 18Nxx 57R90 81R50)},
  MRNUMBER = {4219821},
       DOI = {10.1007/978-3-030-61163-7},
       URL = {https://doi.org/10.1007/978-3-030-61163-7},
}

@article {AFR,
    AUTHOR = {Ayala, David and Francis, John and Rozenblyum, Nick},
     TITLE = {Factorization homology {I}: {H}igher categories},
   JOURNAL = {Adv. Math.},
  FJOURNAL = {Advances in Mathematics},
    VOLUME = {333},
      YEAR = {2018},
     PAGES = {1042--1177},
      ISSN = {0001-8708,1090-2082},
   MRCLASS = {58D29 (18F60 57N80 57R15 57R19 57R56 57S05)},
  MRNUMBER = {3818096},
       DOI = {10.1016/j.aim.2018.05.031},
       URL = {https://doi.org/10.1016/j.aim.2018.05.031},
}

@article{Getal,
    author = "Gaiotto, Davide and Kapustin, Anton and Seiberg, Nathan and Willett, Brian",
    title = "{Generalized Global Symmetries}",
    eprint = "1412.5148",
    archivePrefix = "arXiv",
    primaryClass = "hep-th",
    doi = "10.1007/JHEP02(2015)172",
    journal = "JHEP",
    volume = "02",
    pages = "172",
    year = "2015"
}

@article{CostelloStringMath,
    author = "Costello, Kevin",
    editor = "Donagi, Ron and Douglas, Michael R. and Kamenova, Ljudmila and Rocek, Martin",
    title = "{Integrable lattice models from four-dimensional field theories}",
    eprint = "1308.0370",
    archivePrefix = "arXiv",
    primaryClass = "hep-th",
    doi = "10.1090/pspum/088/01483",
    journal = "Proc. Symp. Pure Math.",
    volume = "88",
    pages = "3--24",
    year = "2014"
}

@article {CosYag,
    AUTHOR = {Costello, Kevin and Yagi, Junya},
     TITLE = {Unification of integrability in supersymmetric gauge theories},
   JOURNAL = {Adv. Theor. Math. Phys.},
  FJOURNAL = {Advances in Theoretical and Mathematical Physics},
    VOLUME = {24},
      YEAR = {2020},
    NUMBER = {8},
     PAGES = {1931--2041},
      ISSN = {1095-0761,1095-0753},
   MRCLASS = {81T60 (82B20 82B23)},
  MRNUMBER = {4320066},
MRREVIEWER = {Georgios\ Linardopoulos},
       DOI = {10.4310/ATMP.2020.v24.n8.a1},
       URL = {https://doi.org/10.4310/ATMP.2020.v24.n8.a1},
}

@article {CEG,
    AUTHOR = {Contreras, Ivan and Elliott, Chris and Gwilliam, Owen},
     TITLE = {Defects via factorization algebras},
   JOURNAL = {Lett. Math. Phys.},
  FJOURNAL = {Letters in Mathematical Physics},
    VOLUME = {113},
      YEAR = {2023},
    NUMBER = {2},
     PAGES = {Paper No. 46, 26},
      ISSN = {0377-9017,1573-0530},
   MRCLASS = {81T70 (14D21 81T20)},
  MRNUMBER = {4576072},
       DOI = {10.1007/s11005-023-01670-2},
       URL = {https://doi.org/10.1007/s11005-023-01670-2},
}

@article {KapDef,
    AUTHOR = {Kapustin, Anton},
     TITLE = {Wilson-'t {H}ooft operators in four-dimensional gauge theories
              and {$S$}-duality},
   JOURNAL = {Phys. Rev. D (3)},
  FJOURNAL = {Physical Review. D. Third Series},
    VOLUME = {74},
      YEAR = {2006},
    NUMBER = {2},
     PAGES = {025005, 14},
      ISSN = {0556-2821},
   MRCLASS = {81T13 (22E70 81T60)},
  MRNUMBER = {2249977},
MRREVIEWER = {Farhang\ Loran},
       DOI = {10.1103/PhysRevD.74.025005},
       URL = {https://doi.org/10.1103/PhysRevD.74.025005},
}

@incollection {Salv,
    AUTHOR = {Salvatore, Paolo},
     TITLE = {Configuration spaces with summable labels},
 BOOKTITLE = {Cohomological methods in homotopy theory ({B}ellaterra, 1998)},
    SERIES = {Progr. Math.},
    VOLUME = {196},
     PAGES = {375--395},
 PUBLISHER = {Birkh\"{a}user, Basel},
      YEAR = {2001},
      ISBN = {3-7643-6588-9},
   MRCLASS = {55R80 (18D50 55P43)},
  MRNUMBER = {1851264},
MRREVIEWER = {Pilar\ C.\ Carrasco},
}

@article {McDuff,
    AUTHOR = {McDuff, Dusa},
     TITLE = {Configuration spaces of positive and negative particles},
   JOURNAL = {Topology},
  FJOURNAL = {Topology. An International Journal of Mathematics},
    VOLUME = {14},
      YEAR = {1975},
     PAGES = {91--107},
      ISSN = {0040-9383},
   MRCLASS = {55D35 (57D25)},
  MRNUMBER = {358766},
MRREVIEWER = {D.\ B.\ Fuchs},
       DOI = {10.1016/0040-9383(75)90038-5},
       URL = {https://doi.org/10.1016/0040-9383(75)90038-5},
}

@article {SegFA,
    AUTHOR = {Segal, Graeme},
     TITLE = {Configuration-spaces and iterated loop-spaces},
   JOURNAL = {Invent. Math.},
  FJOURNAL = {Inventiones Mathematicae},
    VOLUME = {21},
      YEAR = {1973},
     PAGES = {213--221},
      ISSN = {0020-9910,1432-1297},
   MRCLASS = {55D35},
  MRNUMBER = {331377},
MRREVIEWER = {J.\ P.\ May},
       DOI = {10.1007/BF01390197},
       URL = {https://doi.org/10.1007/BF01390197},
}

@incollection {SegCFT,
    AUTHOR = {Segal, Graeme},
     TITLE = {The definition of conformal field theory},
 BOOKTITLE = {Topology, geometry and quantum field theory},
    SERIES = {London Math. Soc. Lecture Note Ser.},
    VOLUME = {308},
     PAGES = {421--577},
 PUBLISHER = {Cambridge Univ. Press, Cambridge},
      YEAR = {2004},
      ISBN = {0-521-54049-6},
   MRCLASS = {81T40 (57R56 81T45)},
  MRNUMBER = {2079383},
MRREVIEWER = {Thomas\ M.\ Fiore and Igor\ K\v{r}\'{\i}\v{z}},
}

@article {AtiTFT,
    AUTHOR = {Atiyah, Michael},
     TITLE = {Topological quantum field theories},
   JOURNAL = {Inst. Hautes \'{E}tudes Sci. Publ. Math.},
  FJOURNAL = {Institut des Hautes \'{E}tudes Scientifiques. Publications
              Math\'{e}matiques},
    NUMBER = {68},
      YEAR = {1988},
     PAGES = {175--186},
      ISSN = {0073-8301,1618-1913},
   MRCLASS = {57R55 (58E15 81E13 81E40)},
  MRNUMBER = {1001453},
MRREVIEWER = {Matthias\ Blau},
}

@book {HaagLQT,
    AUTHOR = {Haag, Rudolf},
     TITLE = {Local quantum physics},
    SERIES = {Texts and Monographs in Physics},
   EDITION = {Second},
      NOTE = {Fields, particles, algebras},
 PUBLISHER = {Springer-Verlag, Berlin},
      YEAR = {1996},
     PAGES = {xvi+390},
      ISBN = {3-540-61451-6; 3-540-61049-9},
   MRCLASS = {81-01 (46L60 46N50 81T05)},
  MRNUMBER = {1405610},
MRREVIEWER = {Alan\ D.\ Sokal},
       DOI = {10.1007/978-3-642-61458-3},
       URL = {https://doi.org/10.1007/978-3-642-61458-3},
}

@article {HaagKast,
    AUTHOR = {Haag, Rudolf and Kastler, Daniel},
     TITLE = {An algebraic approach to quantum field theory},
   JOURNAL = {J. Mathematical Phys.},
  FJOURNAL = {Journal of Mathematical Physics},
    VOLUME = {5},
      YEAR = {1964},
     PAGES = {848--861},
      ISSN = {0022-2488,1089-7658},
   MRCLASS = {81.46},
  MRNUMBER = {165864},
MRREVIEWER = {H.\ Araki},
       DOI = {10.1063/1.1704187},
       URL = {https://doi.org/10.1063/1.1704187},
}

@incollection {Hen,
    AUTHOR = {Henriques, Andr\'{e}},
     TITLE = {Conformal nets are factorization algebras},
 BOOKTITLE = {String-{M}ath 2016},
    SERIES = {Proc. Sympos. Pure Math.},
    VOLUME = {98},
     PAGES = {229--239},
 PUBLISHER = {Amer. Math. Soc., Providence, RI},
      YEAR = {2018},
      ISBN = {978-1-4704-3515-8},
   MRCLASS = {81T40},
  MRNUMBER = {3821755},
MRREVIEWER = {Ali\ Shojaei-Fard},
       DOI = {10.1090/pspum/098/01749},
       URL = {https://doi.org/10.1090/pspum/098/01749},
}

@book {Hua,
    AUTHOR = {Huang, Yi-Zhi},
     TITLE = {Two-dimensional conformal geometry and vertex operator
              algebras},
    SERIES = {Progress in Mathematics},
    VOLUME = {148},
 PUBLISHER = {Birkh\"{a}user Boston, Inc., Boston, MA},
      YEAR = {1997},
     PAGES = {xiv+280},
      ISBN = {0-8176-3829-6},
   MRCLASS = {17B69 (18D10 81-02 81R10 81T40)},
  MRNUMBER = {1448404},
MRREVIEWER = {Mirko\ Primc},
}

@article {WilVir,
    AUTHOR = {Williams, Brian},
     TITLE = {The {V}irasoro vertex algebra and factorization algebras on
              {R}iemann surfaces},
   JOURNAL = {Lett. Math. Phys.},
  FJOURNAL = {Letters in Mathematical Physics},
    VOLUME = {107},
      YEAR = {2017},
    NUMBER = {12},
     PAGES = {2189--2237},
      ISSN = {0377-9017,1573-0530},
   MRCLASS = {81R10 (17B65 17B68 18G55)},
  MRNUMBER = {3719639},
       DOI = {10.1007/s11005-017-0982-7},
       URL = {https://doi.org/10.1007/s11005-017-0982-7},
}

@article {Sharpe,
    AUTHOR = {Sharpe, Eric},
     TITLE = {Notes on generalized global symmetries in {QFT}},
   JOURNAL = {Fortschr. Phys.},
  FJOURNAL = {Fortschritte der Physik. Progress of Physics},
    VOLUME = {63},
      YEAR = {2015},
    NUMBER = {11-12},
     PAGES = {659--682},
      ISSN = {0015-8208,1521-3978},
   MRCLASS = {81R40},
  MRNUMBER = {3422349},
       DOI = {10.1002/prop.201500048},
       URL = {https://doi.org/10.1002/prop.201500048},
}

@article {NT1,
    AUTHOR = {Nest, Ryszard and Tsygan, Boris},
     TITLE = {Algebraic index theorem for families},
   JOURNAL = {Adv. Math.},
  FJOURNAL = {Advances in Mathematics},
    VOLUME = {113},
      YEAR = {1995},
    NUMBER = {2},
     PAGES = {151--205},
      ISSN = {0001-8708,1090-2082},
   MRCLASS = {58G12 (19D55 19K56 46L85 47A53 58G15 58H10)},
  MRNUMBER = {1337107},
MRREVIEWER = {Jonathan\ M.\ Rosenberg},
       DOI = {10.1006/aima.1995.1037},
       URL = {https://doi.org/10.1006/aima.1995.1037},
}

@article {NT2,
    AUTHOR = {Nest, Ryszard and Tsygan, Boris},
     TITLE = {Algebraic index theorem},
   JOURNAL = {Comm. Math. Phys.},
  FJOURNAL = {Communications in Mathematical Physics},
    VOLUME = {172},
      YEAR = {1995},
    NUMBER = {2},
     PAGES = {223--262},
      ISSN = {0010-3616,1432-0916},
   MRCLASS = {58G12 (19D55 19K56 46L85 47A53 58G15 58H10)},
  MRNUMBER = {1350407},
MRREVIEWER = {Jonathan\ M.\ Rosenberg},
       URL = {http://projecteuclid.org/euclid.cmp/1104274104},
}

@article {FR,
    AUTHOR = {Fredenhagen, Klaus and Rejzner, Katarzyna},
     TITLE = {Batalin-{V}ilkovisky formalism in perturbative algebraic
              quantum field theory},
   JOURNAL = {Comm. Math. Phys.},
  FJOURNAL = {Communications in Mathematical Physics},
    VOLUME = {317},
      YEAR = {2013},
    NUMBER = {3},
     PAGES = {697--725},
      ISSN = {0010-3616,1432-0916},
   MRCLASS = {81T15 (81T20 81T70)},
  MRNUMBER = {3009722},
MRREVIEWER = {Ko\ Sanders},
       DOI = {10.1007/s00220-012-1601-1},
       URL = {https://doi.org/10.1007/s00220-012-1601-1},
}

@article {CosLiAnom,
    AUTHOR = {Costello, Kevin and Li, Si},
     TITLE = {Anomaly cancellation in the topological string},
   JOURNAL = {Adv. Theor. Math. Phys.},
  FJOURNAL = {Advances in Theoretical and Mathematical Physics},
    VOLUME = {24},
      YEAR = {2020},
    NUMBER = {7},
     PAGES = {1723--1771},
      ISSN = {1095-0761,1095-0753},
   MRCLASS = {81T50 (81T30)},
  MRNUMBER = {4313234},
MRREVIEWER = {Niccol\`o\ Cribiori},
       DOI = {10.4310/ATMP.2020.v24.n7.a2},
       URL = {https://doi.org/10.4310/ATMP.2020.v24.n7.a2},
}

@unpublished {CosGaiHol,
    AUTHOR = {Costello, Kevin and Gaiotto, Davide},
     TITLE = {Twisted holography},
NOTE = {Available at \url{https://arxiv.org/abs/1812.09257}},
}

@unpublished {CosGaiQ,
    AUTHOR = {Costello, Kevin and Gaiotto, Davide and Yagi, Junya},
     TITLE = {Q-operators are 't {H}ooft lines},
NOTE = {Available at \url{https://arxiv.org/abs/2103.01835}},
}

@unpublished {BudGai,
    AUTHOR = {Budzik, Kasia and Gaiotto, Davide},
     TITLE = {Giant gravitons in twisted holography},
NOTE = {Available at \url{https://arxiv.org/abs/2106.14859}},
}

@unpublished{benini2022quantization,
      title={Quantization of {L}orentzian free {BV} theories: factorization algebra vs algebraic quantum field theory}, 
      author={Marco Benini and Giorgio Musante and Alexander Schenkel},
NOTE = {Available at \url{https://arxiv.org/abs/2212.02546}},
}

@article {CosPaq,
    AUTHOR = {Costello, Kevin and Paquette, Natalie M.},
     TITLE = {Twisted supergravity and {K}oszul duality: a case study in
              {$\rm AdS_3$}},
   JOURNAL = {Comm. Math. Phys.},
  FJOURNAL = {Communications in Mathematical Physics},
    VOLUME = {384},
      YEAR = {2021},
    NUMBER = {1},
     PAGES = {279--339},
      ISSN = {0010-3616,1432-0916},
   MRCLASS = {81T35 (17B81 32G07 81T40 83E50)},
  MRNUMBER = {4252878},
MRREVIEWER = {Eirik\ Eik\ Svanes},
       DOI = {10.1007/s00220-021-04065-3},
       URL = {https://doi.org/10.1007/s00220-021-04065-3},
}

@unpublished{Bru,
      title={Vertex Algebras and {C}ostello-{G}william Factorization Algebras}, 
      author={Daniel Bruegmann},
NOTE = {Available at \url{https://arxiv.org/abs/2012.12214}},
}

@unpublished{ChiHohPin,
      title={Homological Quantum Mechanics}, 
      author={Christoph Chiaffrino and Olaf Hohm and Allison F. Pinto},
NOTE = {Available at \url{https://arxiv.org/abs/2112.11495}},
}

@article{RabThesis,
    author = "Rabinovich, Eugene",
    title = "{Factorization Algebras for Bulk-Boundary Systems}",
NOTE = {Available at \url{https://arxiv.org/abs/2111.01757}},
}

@unpublished{GwiRej2,
      title={The observables of a perturbative algebraic quantum field theory form a factorization algebra}, 
      author={Owen Gwilliam and Kasia Rejzner},
NOTE = {Available at \url{https://arxiv.org/abs/2212.08175}},
}

@unpublished{GuiLi2,
      title={Elliptic Trace Map on Chiral Algebras}, 
      author={Zhengping Gui and Si Li},
NOTE = {Available at \url{https://arxiv.org/abs/2112.14572}},
}

@unpublished{EGW,
      title={Higher deformation quantization of {K}apustin-{W}itten theories}, 
      author={Chris Elliott and Owen Gwilliam and Brian Williams},
NOTE = {Available at \url{https://arxiv.org/abs/2108.13392}},
}

@unpublished{FMT,
      title={Topological symmetry in quantum field theory}, 
      author={Daniel S. Freed and Gregory W. Moore and Constantin Teleman},
NOTE = {Available at \url{https://arxiv.org/abs/2209.07471}},
}

@article {KirTham,
    AUTHOR = {Kirillov, Jr., Alexander and Tham, Ying Hong},
     TITLE = {Factorization homology and 4{D} {TQFT}},
   JOURNAL = {Quantum Topol.},
  FJOURNAL = {Quantum Topology},
    VOLUME = {13},
      YEAR = {2022},
    NUMBER = {1},
     PAGES = {1--54},
      ISSN = {1663-487X,1664-073X},
   MRCLASS = {57R56 (18M20)},
  MRNUMBER = {4404797},
MRREVIEWER = {Haimiao\ Chen},
       DOI = {10.4171/qt/159},
       URL = {https://doi.org/10.4171/qt/159},
}

@incollection {segal2014geometric,
    AUTHOR = {Segal, Graeme},
     TITLE = {A geometric perspective on quantum field theory},
 BOOKTITLE = {Algebraic topology: applications and new directions},
    SERIES = {Contemp. Math.},
    VOLUME = {620},
     PAGES = {281--294},
 PUBLISHER = {Amer. Math. Soc., Providence, RI},
      YEAR = {2014},
      ISBN = {978-0-8218-9474-3},
   MRCLASS = {81Txx},
  MRNUMBER = {3290096},
       DOI = {10.1090/conm/620/12397},
       URL = {https://doi.org/10.1090/conm/620/12397},
}

@incollection {segal2010locality,
    AUTHOR = {Segal, Graeme},
     TITLE = {Locality of holomorphic bundles, and locality in quantum field
              theory},
 BOOKTITLE = {The many facets of geometry},
     PAGES = {164--176},
 PUBLISHER = {Oxford Univ. Press, Oxford},
      YEAR = {2010},
      ISBN = {978-0-19-953492-0},
   MRCLASS = {57R56 (14D23 55P35 58B34 81R60 81T45)},
  MRNUMBER = {2681691},
MRREVIEWER = {Daniel\ S.\ Freed},
       DOI = {10.1093/acprof:oso/9780199534920.003.0009},
       URL = {https://doi.org/10.1093/acprof:oso/9780199534920.003.0009},
}

@article {Kontsevich:2021dmb,
    AUTHOR = {Kontsevich, Maxim and Segal, Graeme},
     TITLE = {Wick rotation and the positivity of energy in quantum field
              theory},
   JOURNAL = {Q. J. Math.},
  FJOURNAL = {The Quarterly Journal of Mathematics},
    VOLUME = {72},
      YEAR = {2021},
    NUMBER = {1-2},
     PAGES = {673--699},
      ISSN = {0033-5606,1464-3847},
   MRCLASS = {81T20 (53C80 81T05)},
  MRNUMBER = {4271398},
MRREVIEWER = {Daniel\ Belti\c{t}\u{a}},
       DOI = {10.1093/qmath/haab027},
       URL = {https://doi.org/10.1093/qmath/haab027},
}

@unpublished{Steff,
      title={Derived $C^\infty$-{G}eometry {I}: {F}oundations}, 
      author={Pelle Steffens},
NOTE = {Available at \url{https://arxiv.org/abs/2304.08671}},
}

@unpublished{AlYo,
      title={Towards non-perturbative BV-theory via derived differential cohesive geometry}, 
      author={Luigi Alfonsi and Charles A. S. Young},
NOTE = {Available at \url{https://arxiv.org/abs/2307.15106}},
}

@unpublished{Car,
      title={Derived differential supergeometry}, 
      author={David Carchedi},
NOTE = {Work in progress},
}

@unpublished{BSEnc,
      title={Operads, homotopy theory and higher categories
in algebraic quantum field theory}, 
      author={Marco Benini and Alexander Schenkel},
NOTE = {Available at \url{https://arxiv.org/abs/2305.03372} and a chapter from the {\it Encyclopedia of Mathematical Physics}, 2nd ed.},
}

\end{document}